\documentclass{article}
\usepackage{amssymb}
\usepackage{amsfonts}
\usepackage{amsmath}

\catcode`\@=11
\def\numberbysection{\@addtoreset{equation}{section}
        \def\theequation{\thesection.\arabic{equation}}}
\numberbysection

\begin{document}

\begin{titlepage}     
\vspace{0.5cm}
\begin{center}
{\Large\bf  $C_{n}^{(1)}$, $D_{n}^{(1)}$ and $A_{2n-1}^{(2)}$ reflection K-matrices}\\
\vspace{1cm}
{\large A. Lima-Santos\footnote{e-mail: dals@df.ufscar.br} \hspace{.5cm} and \hspace{.5cm} R. Malara\footnote{e-mail: malara@df.ufscar.br}} \\
\vspace{1cm}
{\large \em Universidade Federal de S\~ao Carlos, Departamento de F\'{\i}sica \\
Caixa Postal 676, CEP 13569-905~~S\~ao Carlos, Brasil}\\
\end{center}
\vspace{1.2cm}

\begin{abstract}
We investigate the possible regular solutions of the boundary Yang-Baxter equation for the
vertex models associated with the  $C_{n}^{(1)}$, $D_{n}^{(1)}$ and $A_{2n-1}^{(2)}$ affine
Lie algebras. We find three types of solutions with $n$, $n-1$ and $1$ free parameters,
respectively. Special cases and all diagonal solutions are presented separately.

\end{abstract}
\vspace{2cm}
\begin{center}
PACS: 75.10.Jm; 05.90.+m \\
Keywords: Reflection Equations, K-matrices
\end{center}
\vfill
\begin{center}
\small{\today}
\end{center}
\end{titlepage}

\section{Introduction}

The search for integrable models through solutions of the Yang-Baxter
equation \cite{Baxter, KIB, ABR}%
\begin{equation}
\mathcal{R}_{12}(u-v)\mathcal{R}_{13}(u)\mathcal{R}_{23}(v)=\mathcal{R}%
_{23}(v)\mathcal{R}_{13}(u)\mathcal{R}_{12}(u-v)  \label{int.1}
\end{equation}%
has been performed by the quantum group approach in \cite{KR}, where the
problem is reduced to a linear one. Indeed, $\mathcal{R}$-matrices
corresponding to vector representations of all non-exceptional affine Lie
algebras were determined in this way in \cite{Jimbo}.

A similar approach is desirable for finding solutions of the boundary
Yang-Baxter equation \cite{Cherednik, Sklyanin} where the boundary weights
follow from $K$-matrices which satisfy a pair of equations: the reflection
equation%
\begin{equation}
\mathcal{R}_{12}(u-v)K_{1}^{-}(u)\mathcal{R}%
_{12}^{t_{1}t_{2}}(u+v)K_{2}^{-}(v)=K_{2}^{-}(v)\mathcal{R}%
_{12}(u+v)K_{1}^{-}(u)\mathcal{R}_{12}^{t_{1}t_{2}}(u-v),  \label{int.2}
\end{equation}%
and the dual reflection equation%
\begin{equation*}
\mathcal{R}_{12}(-u+v)(K_{1}^{+})^{t_{1}}(u)M_{1}^{-1}\mathcal{R}%
_{12}^{t_{1}t_{2}}(-u-v-2\rho )M_{1}(K_{2}^{+})^{t_{2}}(v)
\end{equation*}%
\begin{equation}
=(K_{2}^{+})^{t_{2}}(v)M_{1}\mathcal{R}_{12}(-u-v-2\rho
)M_{1}^{-1}(K_{1}^{+})^{t_{1}}(u)\mathcal{R}_{12}^{t_{1}t_{2}}(-u+v).
\label{int.3}
\end{equation}%
In this case the duality supplies a relation between $K^{-}$ and $K^{+}$ 
\cite{MN1}:%
\begin{equation}
K^{-}(u):\rightarrow K^{+}(u)=K^{-}(-u-\rho )^{t}M,\qquad M=U^{t}U=M^{t}.
\label{int.4}
\end{equation}%
Here $t_{i}$ denotes transposition in the $i$-\textit{th} space, $\rho $ is
the crossing parameter and $U$ is the crossing matrix, both being specific
to each model \cite{Bazhanov}.

With this goal in mind, the study of boundary quantum groups was initiated
in \cite{MN}. \ These boundary quantum groups have been used to determine A$%
_{1}^{(1)}$ reflection matrices for arbitrary spin \cite{ND}, and the A$%
_{2}^{(2)}$ and some A$_{n}^{(1)}$ reflection matrices were re-derived in 
\cite{Nepo}. More recently, reflection solutions from $\mathcal{R}$-matrices
corresponding to vector representations of Yangians and super-Yangians were
presented in \cite{AACDFR}. \ However, as observed by Nepomechie \cite{Nepo}%
, an independent systematic method of constructing the boundary quantum
group generators is not yet available. In contrast to the bulk case \cite%
{Jimbo}, one can not exploit boundary affine Toda field theory, since
appropriate classical integrable boundary conditions are not yet known \cite%
{BCDR}. Therefore, it is still an open question whether it is possible to
find all solutions of the reflection equations by using quantum group
generators.

Independently, there has been an increasing amount of effort towards the
understanding of two-dimensional integrable theories with boundaries via
solutions of the reflection equation (\ref{int.2}). In field theory,
attention is focused on the boundary $S$-matrix. In statistical mechanics,
the emphasis has been laid on deriving all solutions of (\ref{int.2})
because different $K$-matrices lead to different universality classes of
surface critical behavior \cite{Batch1} and allow the calculation of various
surface critical phenomena, both at and away from criticality \cite{Batch2}.

Although being a hard problem, the direct computation has been used to
derive the solutions of the boundary Yang-Baxter equation (\ref{int.2}) for
given $\mathcal{R}$. For instance, we mention the solutions of the
reflection equations of two-component system \cite{RE1, RE2}, for $19$%
-vertex models \cite{RE3, RE4, RE5, RE6}, for A$_{n-1}^{(1)}$ models \cite%
{RE7, RE8, RE9}, for D$_{n+1}^{(2)}$ models \cite{RE10, RE11}, for A$%
_{2n}^{(2)}$ and B$_{n}^{(1)}$ models \cite{RE12}. Many diagonal solutions
for face and vertex models associated with affine Lie algebras were
presented in \cite{Batch2}. For {\small A-D-E} interaction-round face
models, diagonal and some non-diagonal solutions were presented in \cite%
{Pearce}. \ Reflection matrices for Andrews-Baxter-Forrester models in the 
{\small RSOS/SOS} representation were presented in \cite{Ahn}.

Here we will again touch this issue in order to include the last three
non-exceptional Lie algebras, namely C$_{n}^{(1)}$, D$_{n}^{(1)}$ and A$%
_{2n-1}^{(2)}$, in our classification list \cite{RE9, RE11, RE12}.

We have organized this paper as follows. In Section $2$ we choose the
reflection equations and in Section $3$ and $4$ their solutions are derived.
In Section $5$ all diagonal solutions are presented and in Section $6$ the
special cases are discussed. The last Section is reserved for the conclusion.

\section{Reflection Equations}

Quantum $\mathcal{R}$-matrices for the vertex models associated to the C$%
_{n}^{(1)}$,\ D$_{n}^{(1)}$\ and A$_{2n-1}^{(2)}$ affine Lie algebras as
presented by Jimbo in \cite{Jimbo} have the form 
\begin{eqnarray}
\mathcal{R} &=&a_{1}\sum_{i\neq i^{\prime }}E_{ii}\otimes
E_{ii}+a_{2}\sum_{i\neq j,j^{\prime }}E_{ii}\otimes
E_{jj}+a_{3}\sum_{i<j,i\neq j^{\prime }}E_{ij}\otimes E_{ji}  \notag \\
&&+a_{4}\sum_{i>j,i\neq j^{\prime }}E_{ij}\otimes E_{ji}+\sum_{i,j}a_{ij}\
E_{ij}\otimes E_{i^{\prime }j^{\prime }}  \label{re.1}
\end{eqnarray}%
where $E_{ij}$ denotes the elementary $2n$ by $2n$ matrices ($%
(E_{ij})_{ab}=\delta _{ia}\delta _{jb}$) and the Boltzmann weights with
functional dependence on the spectral parameter $u$ are given by 
\begin{eqnarray}
a_{1}(u) &=&(\mathrm{e}^{u}-q^{2})(\mathrm{e}^{u}-\xi ),\qquad \ a_{2}(u)=q(%
\mathrm{e}^{2u}-1)(\mathrm{e}^{2u}-\xi ),  \notag \\
a_{3}(u) &=&-(q^{2}-1)(\mathrm{e}^{u}-\xi ),\qquad \ a_{4}(u)=\mathrm{e}%
^{u}a_{3}(u)  \label{re.2}
\end{eqnarray}%
and%
\begin{equation}
a_{ij}(u)=\left\{ 
\begin{array}{cc}
(q^{2}\mathrm{e}^{u}-\xi )(\mathrm{e}^{u}-1) & (i=j) \\ 
(q^{2}-1)\ \left( \varepsilon _{i}\varepsilon _{j}\xi q^{\overset{\_}{i}-%
\overset{\_}{j}}(\mathrm{e}^{u}-1)-\delta _{ij^{\prime }}(\mathrm{e}^{u}-\xi
)\right)  & (i<j) \\ 
(q^{2}-1)\mathrm{e}^{u}\left( \varepsilon _{i}\varepsilon _{j}q^{\overset{\_}%
{i}-\overset{\_}{j}}(\mathrm{e}^{u}-1)-\delta _{ij^{\prime }}(\mathrm{e}%
^{u}-\xi )\right)  & (i>j)%
\end{array}%
\right.   \label{re.3}
\end{equation}%
where $q=e^{-2\eta }$ denotes an arbitrary parameter, $\ i^{\prime }=2n+1-i$
\ and $\varepsilon _{i}=1\ (1\leq i\leq n),\ =-1\ (n+1\leq i\leq 2n)$ for C$%
_{n}^{(1)}$ and $\varepsilon _{i}=1$ for A$_{2n-1}^{(2)}$ and D$_{n}^{(1)}$.

Here $\xi =q^{2n+2}$ for C$_{n}^{(1)}$, \ $\xi =q^{2n-2}$ for D$_{n}^{(1)}$
and \ $\xi =-q^{2n}$ for A$_{2n-1}^{(2)}$ and%
\begin{equation}
\overset{\_}{i}=\left\{ 
\begin{array}{c}
i-1/2\ \qquad \qquad \quad \ \ \ (1\leq i\leq n) \\ 
\\ 
i+1/2\ \qquad \quad \ \ \ (n+1\leq i\leq 2n)%
\end{array}%
\right.  \label{re.4}
\end{equation}%
for C$_{n}^{(1)}$, and%
\begin{equation}
\overset{\_}{i}=\left\{ 
\begin{array}{c}
i+1/2\ \qquad \quad \qquad \ \ \ (1\leq i\leq n) \\ 
\\ 
i-1/2\ \qquad \quad \ \ \ (n+1\leq i\leq 2n)%
\end{array}%
\right.  \label{re.5}
\end{equation}%
for A$_{2n-1}^{(2)}$ and \ D$_{n}^{(1)}$.

These $\mathcal{R}$-matrices are regular satisfying {\small PT}-symmetry and
unitarity: 
\begin{equation}
\mathcal{R}(0)=a_{1}(0)\mathcal{P},\qquad \mathcal{R}_{21}(u)=\mathcal{P}%
_{12}\mathcal{R}_{12}(u)\mathcal{P}_{12},\qquad \mathcal{R}_{12}(u)\mathcal{R%
}_{21}(-u)=a_{1}(u)a_{1}(-u)\mathbf{1,}  \label{re.6}
\end{equation}%
where $\mathcal{P}$ is the permutation matrix.\ 

After normalizing the Boltzmann weights by a factor $\sqrt{\xi }\mathrm{e}%
^{u}$, the crossing-unitarity symmetry 
\begin{equation}
\mathcal{R}_{12}(u)=(U\otimes 1)\mathcal{R}_{12}^{t_{2}}(-u-\rho )(U\otimes
1)^{-1},  \label{re.7}
\end{equation}%
holds with the crossing matrices $U$ and crossing parameters $\rho $ given
by 
\begin{equation}
U_{ij}=\delta _{i^{\prime }j}\ q^{(\overset{\_}{i}-\overset{\_}{j}%
)/2},\qquad \rho =-\ln \xi .  \label{re.8}
\end{equation}

Regular solutions of the reflection equation (\ref{int.2}) mean that the
matrix $K^{-}(u)$ in the form 
\begin{equation}
K^{-}=\sum_{i,j=1}^{2n}k_{i,j}(u)\ E_{ij}  \label{re.9}
\end{equation}%
satisfies the condition 
\begin{equation}
k_{i,j}(0)=\delta _{i,j},\quad \qquad i,j=1,2,...,2n.  \label{re.10}
\end{equation}

Substituting (\ref{re.9}) and (\ref{re.1}) into (\ref{int.2}), we will get $%
(2n)^{4}$ functional equations for the $k_{i,j}$ matrix elements, many of
which are dependent. In order to solve them, we shall proceed in the
following way. First we consider the $(i,j)$ component of the matrix
equation (\ref{int.2}). By differentiating it with respect to $v$ and taking 
$v=0$, we get algebraic equations involving the single variable $u$ and $%
4n^{2}$ parameters 
\begin{equation}
\beta _{i,j}=\frac{dk_{i,j}(v)}{dv}|_{v=0}\qquad i,j=1,2,...,2n
\label{re.11}
\end{equation}%
Next we denote these equations by $E[i,j]=0$ and collect them into blocks $%
B[i,j]$, $i=1,...,2n^{2}$ and $j=i,i+1,...,2n^{2}-i$, defined by%
\begin{equation}
B[i,j]=\left\{ 
\begin{array}{c}
E[i,j]=0 \\ 
E[j,i]=0 \\ 
E[4n^{2}+1-i,4n^{2}+1-j]=0 \\ 
E[4n^{2}+1-j,4n^{2}+1-i]=0.%
\end{array}%
\right.  \label{re.12}
\end{equation}%
For a given block $B[i,j]$, the equation $E[4n^{2}+1-i,4n^{2}+1-j]=0$ can be
obtained from the equation $E[i,j]$ by interchanging%
\begin{equation}
k_{i,j}\longleftrightarrow k_{i^{\prime },j^{\prime }},\text{ \ \ \ }\beta
_{i,j}\longleftrightarrow \beta _{i^{\prime },j^{\prime }},\text{ \ \ \ }%
a_{3}\longleftrightarrow a_{4},\text{ \ \ \ }a_{ij}\longleftrightarrow
a_{i^{\prime }j^{\prime }}  \label{re.13}
\end{equation}%
and the equation $E[j,i]=0$ is obtained from the equation $E[i,j]=0$ by
interchanging%
\begin{equation}
k_{i,j}\longleftrightarrow k_{j,i},\text{ \ \ \ }\beta
_{i,j}\longleftrightarrow \beta _{j,i},\text{ \ \ \ }a_{ij}%
\longleftrightarrow a_{j^{\prime }i^{\prime }}.  \label{re.14}
\end{equation}%
Since each $\mathcal{R}$-matrix (\ref{re.1}) satisfies unitarity, P and T
invariances and crossing symmetry, the correspondent $K^{+}$\ matrix is
obtained from the $K^{-}$ matrix using (\ref{int.4}) with the following $M$
matrix%
\begin{equation}
M_{ij}=\delta _{ij\ }q^{2n+1-2\overset{\_}{i}},\qquad i,j=1,2,...,2n.
\label{re.15}
\end{equation}

\section{General Solutions}

The challenge at this point is to find the full set of solutions of the
reflection equations (\ref{int.2}) and (\ref{int.3}) for $\mathcal{R}$%
-matrices associated to C$_{n}^{(1)}$, D$_{n}^{(1)}$ and A$_{2n-1}$ affine
Lie algebras. We will start by looking at $K$-matrix solutions with all
entries different from zero which will be named general solutions.

\subsection{Non-diagonal matrix elements}

Analyzing the reflection equation (\ref{int.2})\ with the $\mathcal{R}$%
-matrix (\ref{re.1}), we can see that\ the simplest functional equations are
those involving only two matrix elements of the type $k_{i,i^{^{\prime }}}$\
(secondary diagonal). They belong to the blocks $%
B[1,2n+3],B[1,4n+5],B[1,6n+7],...$, and we choose to express their solutions
in terms of the element $k_{1,2n}$with $\beta _{1,2n}\neq 0$:%
\begin{equation}
k_{i,i^{^{\prime }}}=\left( \frac{\beta _{i,i^{^{\prime }}}}{\beta _{1,2n}}%
\right) k_{1,2n}.  \label{gs.1}
\end{equation}%
Next, we look at the last blocks of the collection $\{B[1,j]\}$. Here we can
write the matrix elements of the first row $k_{1,j}\ (j\neq 1,2n)$ in terms
of the element $k_{1,2n}$ and their transpose in terms of the element $%
k_{2n,1}$. From the last blocks of the collection $\{B[2n+3,j]\}$, the
matrix elements of the second row $k_{2,j}\ (j\neq 2,2n)$ are expressed in
terms of $k_{2,2n-1}$ \ and their transpose in terms of $k_{2n-1,2}$.
Following this procedure with the collections $\{B[4n+5,j]\},\{B[6n+7,j]%
\},...,$ we will be able to write all non-diagonal matrix elements as:%
\begin{equation}
k_{i,j}=\left( \frac{a_{1}a_{11}-a_{2}^{2}}{%
a_{3}a_{4}a_{11}^{2}-a_{2}^{2}a_{12}a_{21}}\right) \left( \beta
_{i,j}a_{3}a_{11}-\beta _{j^{^{\prime }},i^{^{\prime }}}a_{2}a_{ij^{^{\prime
}}}\right) \frac{k_{1,2n}}{\beta _{1,2n}}\qquad (j<i^{^{\prime }})
\label{gs.2}
\end{equation}%
and%
\begin{equation}
k_{i,j}=\left( \frac{a_{1}a_{11}-a_{2}^{2}}{%
a_{3}a_{4}a_{11}^{2}-a_{2}^{2}a_{12}a_{21}}\right) \left( \beta
_{i,j}a_{4}a_{11}-\beta _{j^{^{\prime }},i^{^{\prime }}}a_{2}a_{ij^{^{\prime
}}}\right) \frac{k_{1,2n}}{\beta _{1,2n}}\qquad (j>i^{^{\prime }})
\label{gs.3}
\end{equation}%
where we have used \ (\ref{gs.1}) and the identities%
\begin{equation}
a_{ij}=a_{j^{^{\prime }}i^{^{\prime }}}\qquad \mathrm{and}\qquad
a_{1j}a_{j1}=a_{12}a_{21}\qquad (j\neq 1).  \label{gs.4}
\end{equation}

Taking into account the Boltzmann weights of each model, we substitute these
expressions\ in the remaining \ reflection equations and look at for those
without diagonal entries $k_{i,i}$, in order to fix some parameters $\beta
_{i,j}$ ($i\neq j$). For instance, from the blocks $B[i,i]$ one can see that
their equations are solved by the relations%
\begin{equation}
\beta _{i,j}k_{j,i}=\beta _{j,i}k_{i,j}\qquad \Longrightarrow \qquad \beta
_{i,j}\beta _{j^{^{\prime }},i^{^{\prime }}}=\beta _{j,i}\beta _{i^{^{\prime
}},j^{^{\prime }}}.  \label{gs.5}
\end{equation}%
\ 

From the equations $E[2,j]$ and $E[j,2]$ we can find two possibilities to
express the parameters for the matrix elements below the secondary diagonal (%
$\beta _{i,j}$ with $j>i^{^{\prime }}$) in terms of those above of the
secondary diagonal 
\begin{equation}
\beta _{i,j}=\left\{ 
\begin{array}{c}
\pm \theta _{i}\frac{1}{\sqrt{\xi }}q^{\frac{1}{2}(\overset{\_}{i}-\overset{%
\_}{i^{^{\prime }}})+j-n-1}\ \beta _{j^{^{\prime }},i^{^{\prime }}}\qquad 
\mathrm{for}\qquad j>n+1 \\ 
\\ 
\pm \theta _{j}\frac{1}{\sqrt{\xi }}q^{\frac{1}{2}(\overset{\_}{j}-\overset{%
\_}{j^{^{\prime }}})+i-n-1}\ \beta _{j^{^{\prime }},i^{^{\prime }}}\qquad 
\mathrm{for}\qquad j\leq n+1%
\end{array}%
\right.  \label{gs.6}
\end{equation}%
where $\theta _{i}=q\varepsilon _{i}$ for C$_{n}^{(1)}$ , $\theta _{i}=1$
for A$_{2n-1}^{(2)}$ and D$_{n}^{(1)}$.

These relations simplify the expressions for the non-diagonal matrix
elements (\ref{gs.2}) and (\ref{gs.3}):%
\begin{equation}
k_{i,j}(u)=\left\{ 
\begin{array}{c}
\beta _{i,j}\mathcal{G}^{(\pm )}(u),\qquad \qquad \quad \ \ \ \ \ \ \ \
(j<i^{^{\prime }}) \\ 
\\ 
\beta _{i,i^{^{\prime }}}\left( \frac{q\mathrm{e}^{u}\pm \sqrt{\xi }}{q\pm 
\sqrt{\xi }}\right) \mathcal{G}^{(\pm )}(u)\qquad \quad \qquad
(j=i^{^{\prime }}) \\ 
\\ 
\beta _{i,j}\mathrm{e}^{u}\mathcal{G}^{(\pm )}(u)\qquad \qquad \ \ \ \ \ \ \
\ \ \ (j>i^{^{\prime }})%
\end{array}%
\right. ,  \label{gs.7}
\end{equation}%
where $\mathcal{G}^{(\pm )}(u)$ are defined conveniently by a normalization
of $k_{1,2n}(u):$ 
\begin{equation}
\mathcal{G}^{(\pm )}(u)=\frac{1}{\beta _{1,2n}}\left( \frac{q\pm \sqrt{\xi }%
}{q\mathrm{e}^{u}\pm \sqrt{\xi }}\right) k_{1,2n}(u).  \label{gs.8}
\end{equation}

Now, we substitute these expressions in the remaining equations $E[2,j]$ and 
$E[j,2]$, and look at the equations of the type 
\begin{equation}
F(u)\mathcal{G}^{(\pm )}(u)=0  \label{gs.9}
\end{equation}%
where $F(u)=\sum_{k}f_{k}(\left\{ \beta _{ij}\right\} )\mathrm{e}^{ku}$. The
constraint equations $f_{k}(\left\{ \beta _{ij}\right\} )\equiv 0,\forall k$%
, can be solved in terms of the $2n$ parameters. Of course the expressions
for $k_{ij}$ will depend on our choice of these parameters. The choice $%
\beta _{12}$, $\beta _{13}$, ..., $\beta _{1,2n}$ and $\beta _{21}$ is the
most appropriate for our purpose.

Taking into account all fixed\ parameters in terms of these $2n$ parameters,
we can rewrite the $k_{ij}$ $(i\neq j)$ matrix elements of $K^{-}(u)$ \ for $%
n>2$ in the following way:

The secondary diagonal has the following entries 
\begin{equation}
k_{i,i^{\prime }}(u)=\varepsilon _{i}\frac{\beta _{1,i^{\prime }}^{2}}{\beta
_{1,2n}}\frac{q^{\overset{\_}{i}-\overset{\_}{1}}}{\xi }\frac{\left( q\pm 
\sqrt{\xi }\right) }{\left( q+1\right) ^{2}}\left( q\mathrm{e}^{u}\pm \sqrt{%
\xi }\right) \mathcal{G}^{(\pm )}(u)\qquad (i\neq 1,2n)  \label{gs.10}
\end{equation}%
and 
\begin{equation}
k_{2n,1}(u)=\varepsilon _{2n}\beta _{1,2n}\frac{\beta _{21}^{2}}{\beta
_{1,2n-1}^{2}}q^{\overset{\_}{(2n}-\overset{\_}{1})-2}\left( \frac{q\mathrm{e%
}^{u}\pm \sqrt{\xi }}{q\pm \sqrt{\xi }}\right) \mathcal{G}^{(\pm )}(u).
\label{gs.11}
\end{equation}%
The first row (column) and the last row (column) with $i,j\neq 1,2n$ are:%
\begin{equation}
k_{1,j}(u)=\beta _{1,j}\ \mathcal{G}^{(\pm )}(u),\qquad
k_{i,1}(u)=\varepsilon _{i}\beta _{21}\frac{\beta _{1,i^{^{\prime }}}}{\beta
_{1,2n-1}}q^{\overset{\_}{i}-\overset{\_}{2}}\mathcal{G}^{(\pm )}(u),
\label{gs.12}
\end{equation}%
\begin{equation}
k_{i,2n}(u)=\varepsilon _{i}\beta _{1,i^{\prime }}\frac{q^{\overset{\_}{i}-%
\overset{\_}{1}}}{\pm \sqrt{\xi }}\mathrm{e}^{u}\mathcal{G}^{(\pm
)}(u),\qquad k_{2n,j}(u)=\beta _{21}\frac{\beta _{1,j}}{\beta _{1,2n}}\frac{%
q^{\overset{\_}{2n}-\overset{\_}{2}}}{\pm \sqrt{\xi }}\mathrm{e}^{u}\mathcal{%
G}^{(\pm )}(u),  \label{gs.13}
\end{equation}%
and the remaining non-diagonal matrix elements are given by%
\begin{eqnarray}
k_{ij}(u) &=&\varepsilon _{i}\beta _{1,j}\frac{\beta _{1,i^{\prime }}}{\beta
_{1,2n}}\frac{q^{\overset{\_}{i}-\overset{\_}{1}}}{\pm \sqrt{\xi }}\left( 
\frac{q\pm \sqrt{\xi }}{q+1}\right) \mathcal{G}^{(\pm )}(u),\qquad \ \ \ \ \
\ \ \ \ (j<i^{\prime })  \label{gs.14} \\
k_{ij}(u) &=&\varepsilon _{i}\beta _{1,j}\frac{\beta _{1,i^{\prime }}}{\beta
_{1,2n}}\frac{q^{\overset{\_}{i}-\overset{\_}{1}}}{\xi }\left( \frac{q\pm 
\sqrt{\xi }}{q+1}\right) \mathrm{e}^{u}\mathcal{G}^{(\pm )}(u).\quad \ \ \ \
\ \ \ \ \ \ \ (j>i^{\prime })  \label{gs.15}
\end{eqnarray}%
These relations solve all reflection equations without diagonal entries $%
k_{ii}(u)$, $i=1,2,...,2n$.

\subsection{Diagonal matrix elements}

At this point the remaining reflection equations involve $2n$ diagonal
matrix elements $k_{ii}(u)$, the function $\mathcal{G}^{(\pm )}(u)$ and $4n$
parameters. From the equations $E[1,2]$ and $E[1,2n+1]$ we can find $%
k_{11}(u)$ and $k_{22}(u)$ and from the equations $E[4n^{2},4n^{2}-1]$ and $%
E[4n^{2},4n^{2}+2n]$ we can find $k_{2n,2n}(u)$ and $k_{2n-1,2n-1}(u).$
Next, from the equations $E[2,j]$, $j=3,4,...,2n-2$ we can find the $%
k_{jj}(u)$ matrix elements.

The expressions obtained in this way for the diagonal elements are too
large. However, after finding the following $n-1$ non-diagonal parameters 
\begin{equation}
\beta _{21}=-\frac{q}{\xi }\left( \frac{q\pm \sqrt{\xi }}{q+1}\right) ^{2}%
\frac{\beta _{12}\beta _{1,2n-1}^{2}}{\beta _{1,2n}^{2}},\qquad \beta
_{1,j}=(-1)^{n+j}\frac{\beta _{1,n}\beta _{1,n+1}}{\beta _{1,2n+1-j}},\qquad
j=2,3,...,n-1  \label{gs.16}
\end{equation}%
they are related with $k_{11}(u)$ in a very simple way :

\begin{equation}
k_{i,i}(u)=k_{11}(u)+(\beta _{i,i}-\beta _{11})\mathcal{G}^{(\pm )}(u)\qquad
\qquad \qquad \qquad \qquad \qquad (2\leq i\leq n)  \label{gs.17}
\end{equation}%
\begin{equation}
k_{n+1,n+1}(u)=k_{n,n}(u)+\left( \beta _{n+1,n+1}-\beta _{n,n}\right) 
\mathrm{e}^{u}\mathcal{G}^{(\pm )}(u)+\mathcal{H}^{(\pm )}(u)  \label{gs.18}
\end{equation}%
\begin{equation}
k_{i,i}(u)=k_{n+1,n+1}(u)+\left( \beta _{i,i}-\beta _{n+1,n+1}\right) 
\mathrm{e}^{u}\mathcal{G}^{(\pm )}(u)\qquad \qquad \qquad (n+2\leq i\leq 2n)
\label{gs.19}
\end{equation}%
where%
\begin{equation}
\mathcal{H}^{(+)}(u)=-\Delta _{n}^{(\pm )}(-q)^{n-1}\frac{(\theta
_{n+1}^{2}-\varepsilon _{n+1})}{\left( q+1\right) ^{2}}\left( \mathrm{e}%
^{u}-1\right) \mathcal{G}^{(\pm )}(u)  \label{gs.20}
\end{equation}%
with%
\begin{equation}
\Delta _{n}^{(\pm )}=(-1)^{n}\frac{\beta _{1,n}\beta _{1,n+1}}{\beta _{1,2n}}%
\left( \frac{q\pm \sqrt{\xi }}{\pm \sqrt{\xi }}\right)  \label{gs.21}
\end{equation}%
Here we note that $\mathcal{H}^{(\pm )}(u)=0$ for A$_{2n-1}^{(2)}$ and D$%
_{n}^{(1)}$ models.

An important simplification occurs when we consider the equation $E[2n+1,4n]$
separately. This equation gives an additional relation between $k_{2n,2n}(u)$
and $k_{11}(u)$: 
\begin{equation}
k_{2n,2n}(u)=\mathrm{e}^{2u}k_{11}(u)+(\beta _{2n,2n}-\beta _{11}-2)\mathrm{e%
}^{u}\left( \frac{q\mathrm{e}^{u}\pm \sqrt{\xi }}{q\pm \sqrt{\xi }}\right) 
\mathcal{G}^{(\pm )}(u).  \label{gs.22}
\end{equation}%
Taking $i=2n$ into (\ref{gs.19}) and comparing with (\ref{gs.22}) we can
find the following expression for $k_{11}(u)$:%
\begin{eqnarray}
k_{11}(u) &=&\frac{\mathcal{H}^{(\pm )}(u)}{\mathrm{e}^{2u}-1}+\left\{ \beta
_{n,n}-\beta _{11}+(\beta _{2n,2n}-\beta _{n,n})\mathrm{e}^{u}\right.  \notag
\\
&&\left. -(\beta _{2n,2n}-\beta _{11}-2)\mathrm{e}^{u}\left( \frac{q\mathrm{e%
}^{u}\pm \sqrt{\xi }}{q\pm \sqrt{\xi }}\right) \right\} \frac{\mathcal{G}%
^{(\pm )}(u)}{\mathrm{e}^{2u}-1}  \label{gs.23}
\end{eqnarray}

$\allowbreak $Substituting these expressions into the reflection equations
we will get constraint equations to fix some of the $3n-1$ remaining
parameters. In order to do this we recall the equations $E[2,2n+j]$ to find $%
\beta _{jj}$, $j=3,4,...,2n-2$ in terms of $\beta _{22}$. Then the $\beta
_{22}$ is given by the equations $E[2,2n+1]$. After performing this we can
use the equation $E[2,2n-1]$ to find $\beta _{2n-1,2n-1}$ and $\beta
_{2n,2n} $. These parameters can be written in terms of $\beta _{11},\beta
_{1,n},\beta _{1,n+1}$ and $\beta _{1,2n}$ in the following way:%
\begin{equation}
\beta _{i,i}=\beta _{11}+\Delta _{n}^{(\pm )}\sum_{j=0}^{i-2}(-q)^{j}\ \ \ \
\ \ \ \ \ \ \ \ \ \ \ \ \ (1<i\leq n)  \label{gs.24}
\end{equation}%
\begin{equation}
\beta _{n+1,n+1}=\beta _{11}+\Delta _{n}^{(\pm )}\left[ \frac{1-(-q)^{n-1}}{%
q+1}+(-q)^{n-1}\frac{\left( \theta _{n+1}^{2}-\varepsilon _{n+1}\right)
(q\pm \sqrt{\xi })}{\pm \sqrt{\xi }(q+1)^{2}}\right]  \label{gs.25}
\end{equation}%
\begin{equation}
\beta _{i,i}=\beta _{n+1,n+1}+\Delta _{n}^{(\pm )}\left[ \frac{\theta
_{n+1}^{2}\varepsilon _{n+1}}{\pm \sqrt{\xi }}\sum_{j=n-1}^{i-3}(-q)^{j}\ %
\right] \ \ \ \ \ \ \ \ \ \ \ \ \ \ \ \ (n+2<i\leq 2n-1)  \label{gs.26}
\end{equation}%
\begin{equation}
\beta _{2n,2n}=\beta _{11}+2+\Delta _{n}^{(\pm )}\frac{(q\pm \sqrt{\xi })}{%
\xi }\frac{(\xi -\varepsilon _{2n}q^{\overset{\_}{2n}-\overset{\_}{1}})}{%
\left( q+1\right) ^{2}}  \label{gs.27}
\end{equation}%
Note that $\beta _{n+1,n+1}=\beta _{n,n}$ or that $k_{n+1,n+1}(u)=k_{n,n}(u)$
for A$_{2n-1}^{(2)}$ and D$_{n}^{(1)}$.

Finally, we can use, for instance, the equation $E[2,4n]$ to fix $\beta
_{1,n}$:%
\begin{equation}
\beta _{1,n}=(-1)^{n}\frac{2\xi (\pm \sqrt{\xi })(q+1)^{2}}{(1\mp \sqrt{\xi }%
)\left[ \theta _{1}q^{n-1}\mp (-1)^{n}\sqrt{\xi }\right] \left[ \theta
_{n+1}q^{n}\mp (-1)^{n}\sqrt{\xi }\right] (q\pm \sqrt{\xi })}\frac{\beta
_{1,2n}}{\beta _{1,n+1}}.  \label{gs.28}
\end{equation}

At this point it was possible to treat these solutions simultaneously. But
now we needed separate them in order to take into account the existence of
the amplitude $k_{1,n}(u)$ for each model:

\begin{itemize}
\item For A$_{2n-1}^{(2)}$ models $\xi =-q^{2n}$ and $\theta
_{k}=\varepsilon _{k}=1,\forall k$. For these models there are no
restrictions in (\ref{gs.28}). It follows that the solution with $\mathcal{G}%
^{(+)}(u)$ (up sign) is related to the solution with $\mathcal{G}^{(-)}(u)$
(down sign) by complex conjugation.

\item For D$_{n}^{(1)}$ models \ we have $\xi =q^{2n-2}$, $\theta
_{k}=\varepsilon _{k}=1,\forall k$. It means that the factors $\left[
q^{n-1}\mp (-1)^{n}\sqrt{\xi }\right] $ are different from zero for the
solution with $\mathcal{G}^{(+)}(u)$ only if $n$ is odd and for the solution
with $\mathcal{G}^{(-)}(u)$ if $n$ is even.

\item For C$_{n}^{(1)}$ models $\xi =q^{2n+2}$ and $\theta _{1}=-\theta
_{n+1}=q$. \ In this case the factors $\left[ -q^{n+1}\mp (-1)^{n}\sqrt{\xi }%
\right] $ are different from zero for the solution with $\mathcal{G}%
^{(+)}(u) $ if $n$ is even and for the solution with $\mathcal{G}^{(-)}(u)$
if $n$ is odd.
\end{itemize}

Therefore, we have find\ two general solution for the A$_{2n-1}^{(2)}$
models and only one general solution for C$_{n}^{(1)}$ and D$_{n}^{(1)}$
models.

Substituting all fixed parameters into (\ref{gs.23}) we will find the
following expressions for the amplitude $k_{11}(u)$: 
\begin{eqnarray}
k_{11}(u) &=&\frac{2\mathrm{e}^{u}\mathcal{G}^{(\pm )}(u)}{\mathrm{e}^{2u}-1}
\notag \\
&&-\frac{2\mathcal{G}^{(\pm )}(u)}{\mathrm{e}^{u}+1}\left\{ \frac{\xi \left(
1+q-(-q)^{n-1}+(-q)^{n}\right) +q\mathrm{e}^{u}(\xi -q^{2n-2})}{(1\mp \sqrt{%
\xi })\left[ q^{n-1}\mp (-1)^{n}\sqrt{\xi }\right] \left[ q^{n}\mp (-1)^{n}%
\sqrt{\xi }\right] }\right\}  \label{gs.30}
\end{eqnarray}%
for A$_{2n-1}^{(2)}$ and D$_{n}^{(1)}$ models and%
\begin{eqnarray}
k_{11}(u) &=&\frac{2\mathrm{e}^{u}\mathcal{G}^{(\pm )}(u)}{\mathrm{e}^{2u}-1}
\notag \\
&&+\frac{2\mathcal{G}^{(\pm )}(u)}{\mathrm{e}^{u}+1}\left\{ \frac{\xi \left(
1+q+(-q)^{n}+(-q)^{n+1}\right) +q\mathrm{e}^{u}(\xi +q^{2n})}{(1\mp \sqrt{%
\xi })\left[ q^{n}\mp (-1)^{n}\sqrt{\xi }\right] \left[ q^{n+1}\pm (-1)^{n}%
\sqrt{\xi }\right] }\right\}  \label{gs.31}
\end{eqnarray}%
for C$_{n}^{(1)}$ models.

From (\ref{gs.30}) we can see that $k_{11}(u)$ are quite simple for the D$%
_{n}^{(1)}$models 
\begin{equation}
k_{11}(u)=\frac{\mathcal{G}^{(+)}(u)}{\mathrm{e}^{u}-1}\quad (n-\mathrm{odd}%
),\qquad k_{11}(u)=\frac{\mathcal{G}^{(-)}(u)}{\mathrm{e}^{u}-1}\quad (n-%
\mathrm{even}).  \label{gs.32}
\end{equation}%
Moreover, substituting (\ref{gs.31}) into (\ref{gs.17}--\ref{gs.19}) we will
find a simple relation between the diagonal matrix elements for C$_{n}^{(1)}$
models: 
\begin{equation}
k_{n+i,n+i}(u)=\mathrm{e}^{u}k_{i,i}(u)\Longrightarrow \beta
_{n+i,n+i}=\beta _{i,i}+1.\qquad 1\leq i\leq n  \label{gs.33}
\end{equation}

Now, let us summarize these results: First, we have from (\ref{gs.10}) to (%
\ref{gs.15}) all non-diagonal matrix elements after substituting the $n$
fixed non-diagonal parameters $\beta _{21}$ and $\beta _{1,j}\ (j=2,...,n)$
given by (\ref{gs.16}) and (\ref{gs.28}).

Second, the diagonal matrix elements are obtained by using (\ref{gs.17}), (%
\ref{gs.18}) and (\ref{gs.19}) with $k_{11}(u)$ given by (\ref{gs.30}) for A$%
_{2n-1}^{(2)}$ and D$_{n}^{(1)}$ models and by (\ref{gs.31}) for C$%
_{n}^{(1)} $ models and substituting the diagonal parameters given by (\ref%
{gs.24}), (\ref{gs.25}), (\ref{gs.26}) and (\ref{gs.27}). This sequence
results in two solutions with $n+1$ parameters $\beta _{1,n+1},\beta
_{1,n+2},\ldots ,\beta _{1,2n}$ and $\beta _{11}$for A$_{2n-1}^{(2)}$ models
and one solution for C$_{n}^{(1)}$ and D$_{n}^{(1)}$ models.

The number of free parameters in these general solutions is $n$ because we
still have to use the regular condition (\ref{re.10}), which will fix the
parameter $\beta _{11}$.

Here we note that $n=1$ and $n=2$ are special cases and will be treated in
the section $6$.

\section{Reduced Solutions}

In the previous section we have considered reflection $K$-matrices with all
entries different from zero. In particular, for C$_{n}^{(1)}$ and D$%
_{n}^{(1)}$ models these matrices depend on the parity of $n$: one general
solution with $\mathcal{G}^{(+)}(u)$ for $n$ odd and one general solution
with $\mathcal{G}^{(-)}(u)$ for $n$ even for D$_{n}^{(1)}$ models and the
opposite for the C$_{n}^{(1)}$ models. However, we have found that there
exist solutions with $\mathcal{G}^{(+)}(u)$ for $n$ even and solutions with $%
\mathcal{G}^{(-)}(u)$ for $n$ odd for the D$_{n}^{(1)}$ models and the
opposite for the C$_{n}^{(1)}$ models provided that we allow some matrix
elements to be equal to zero. In order to show this we recall (\ref{gs.1}--%
\ref{gs.3}) to see that the vanishing of the secondary diagonal element $%
k_{n+1,n}(u)$ implies that $k_{i,n}(u)=0$ ($i\neq n$) and $k_{n+1,j}(u)=0$ ($%
j\neq n+1$). Therefore we can consider the case $k_{n+1,n}(u)=0$ which
implies that $k_{n,n+1}(u)=k_{i,n+1}(u)=k_{n,j}(u)=0$ for these models. It
means that we are working out with $K$-matrices that contain $2(4n-3)$ null
entries. In particular, with $k_{1,n}(u)=0$.

The non-diagonal matrix elements for these solutions can be directly
obtained\ from (\ref{gs.10}--\ref{gs.15}) taking into account the limits $%
\beta _{1,n}\rightarrow 0$ and $\beta _{1,n+1}\rightarrow 0$.

The secondary diagonal is 
\begin{eqnarray}
k_{i,i^{\prime }}(u) &=&\varepsilon _{i}\frac{\beta _{1,i^{\prime }}^{2}}{%
\beta _{1,2n}}\frac{q^{\overset{\_}{i}-\overset{\_}{1}}}{\xi }\frac{\left(
q\pm \sqrt{\xi }\right) }{\left( q+1\right) ^{2}}\left( q\mathrm{e}^{u}\pm 
\sqrt{\xi }\right) \mathcal{G}^{(\pm )}(u),\qquad (i\neq 1,n,n+1,2n)
\label{rs.1} \\
k_{n,n+1}(u) &=&k_{n+1,n}(u)=0,  \label{rs.2} \\
k_{2n,1}(u) &=&\varepsilon _{2n}\beta _{1,2n}\frac{\beta _{21}^{2}}{\beta
_{1,2n-1}^{2}}q^{\overset{\_}{(2n}-\overset{\_}{1})-2}\left( \frac{q\mathrm{e%
}^{u}\pm \sqrt{\xi }}{q\pm \sqrt{\xi }}\right) \mathcal{G}^{(\pm )}(u).
\label{rs.3}
\end{eqnarray}%
The boundary rows and columns are 
\begin{eqnarray}
k_{1,j}(u) &=&\beta _{1,j}\ \mathcal{G}^{(\pm )}(u),\qquad
k_{i,1}(u)=\varepsilon _{i}\beta _{21}\frac{\beta _{1,i^{^{\prime }}}}{\beta
_{1,2n-1}}q^{\overset{\_}{i}-\overset{\_}{2}}\mathcal{G}^{(\pm )}(u),
\label{rs.4} \\
k_{1,n}(u) &=&k_{1,n+1}(u)=k_{n,1}(u)=k_{n+1,1}(u)=0,  \label{rs.5}
\end{eqnarray}%
\begin{eqnarray}
k_{i,2n}(u) &=&\varepsilon _{i}\beta _{1,i^{\prime }}\frac{q^{\overset{\_}{i}%
-\overset{\_}{1}}}{\pm \sqrt{\xi }}\mathrm{e}^{u}\mathcal{G}^{(\pm
)}(u),\qquad k_{2n,,j}(u)=\beta _{21}\frac{\beta _{1,j}}{\beta _{1,2n}}\frac{%
q^{\overset{\_}{2n}-\overset{\_}{2}}}{\pm \sqrt{\xi }}\mathrm{e}^{u}\mathcal{%
G}^{(\pm )}(u),  \label{rs.6} \\
k_{n,2n}(u) &=&k_{n+1,2n}(u)=k_{2n,n}(u)=k_{2n,n+1}(u)=0,  \label{rs.7}
\end{eqnarray}%
and the remaining non-diagonal matrix elements are 
\begin{eqnarray}
k_{ij}(u) &=&\varepsilon _{i}\beta _{1,j}\frac{\beta _{1,i^{\prime }}}{\beta
_{1,2n}}\frac{q^{\overset{\_}{i}-\overset{\_}{1}}}{\pm \sqrt{\xi }}\left( 
\frac{q\pm \sqrt{\xi }}{q+1}\right) \mathcal{G}^{(\pm )}(u),\qquad \ \ \ \ \
\ \ \ \ (j<i^{\prime })  \label{rs.8} \\
k_{ij}(u) &=&\varepsilon _{i}\beta _{1,j}\frac{\beta _{1,i^{\prime }}}{\beta
_{1,2n}}\frac{q^{\overset{\_}{i}-\overset{\_}{1}}}{\xi }\left( \frac{q\pm 
\sqrt{\xi }}{q+1}\right) \mathrm{e}^{u}\mathcal{G}^{(\pm )}(u),\quad \ \ \ \
\ \ \ \ \ \ \ (j>i^{\prime })  \label{rs.9} \\
k_{in}(u) &=&k_{i,n+1}(u)=k_{n,j}(u)=k_{n+1,j}(u)=0.  \label{rs.10}
\end{eqnarray}%
In order to find the corresponding diagonal elements we follow the steps
presented previously, but now using the equations $E[n,2n(n-1)+2]$ and $%
E[n+1,2n^{2}+2]$ to find $k_{n,n}(u)$ and $k_{n+1,n+1}(u)$, respectively.
Next we find the following $n-2$ non-diagonal parameters 
\begin{equation}
\beta _{21}=-\frac{q}{\xi }\left( \frac{q\pm \sqrt{\xi }}{q+1}\right) ^{2}%
\frac{\beta _{12}\beta _{1,2n-1}^{2}}{\beta _{1,2n}^{2}},\qquad \beta
_{1,j}=(-1)^{n-1+j}\frac{\beta _{1,n-1}\beta _{1,n+2}}{\beta _{1,2n+1-j}}%
,\qquad j=2,3,...,n-2  \label{rs.11}
\end{equation}%
with $n>3$, they are also related with $k_{11}(u)$ in a very simple way :

\begin{equation}
k_{ii}(u)=k_{11}(u)+(\beta _{i,i}-\beta _{11})\mathcal{G}^{(\pm )}(u),\qquad
1<i\leq n-1  \label{rs.12}
\end{equation}%
\begin{equation}
k_{n,n}(u)=k_{n+1,n+1}(u)=k_{n-1,n-1}(u)+(\beta _{n,n}-\beta _{n-1,n-1})%
\mathrm{e}^{u}\mathcal{G}^{(\pm )}(u)+\mathcal{F}_{n-1}^{(\pm )}(u)
\label{rs.13}
\end{equation}%
\begin{equation}
k_{ii}(u)=k_{n,n}(u)+(\beta _{i,i}-\beta _{n,n})\mathrm{e}^{u}\mathcal{G}%
^{(\pm )}(u)-\varepsilon _{i}\theta _{i}^{2}q^{2}\mathcal{F}_{n-1}^{(\pm
)}(u),\qquad n+2\leq i\leq 2n  \label{rs.14}
\end{equation}%
where%
\begin{equation}
\mathcal{F}_{n-1}^{(\pm )}(u)=-(-q)^{n-2}\Delta _{n-1}^{(\pm )}\frac{(%
\mathrm{e}^{u}-1)}{(q+1)^{2}}\mathcal{G}^{(\pm )}(u)  \label{rs.15}
\end{equation}%
with%
\begin{equation}
\Delta _{n-1}^{(\pm )}=(-1)^{n-1}\frac{\beta _{1,n-1}\beta _{1,n+2}}{\beta
_{1,2n}}\left( \frac{q\pm \sqrt{\xi }}{\pm \sqrt{\xi }}\right) .
\label{rs.16}
\end{equation}%
Here we note that $\Delta _{n-1}^{(\pm )}$ can be understood as a limit $%
(\beta _{1,n}\rightarrow -\beta _{1,n-1};\beta _{1,n+1}\rightarrow \beta
_{1,n+2})$ of $\Delta _{n}^{(\pm )}$ given by (\ref{gs.21}).

Again, the equation $E[2n+1,4n]$ gives us another relation between $%
k_{2n,2n}(u)$ and $k_{11}(u)$: 
\begin{equation}
k_{2n,2n}(u)=\mathrm{e}^{2u}k_{11}(u)+(\beta _{2n,2n}-\beta _{11}-2)\mathrm{e%
}^{u}\left( \frac{q\mathrm{e}^{u}\pm \sqrt{\xi }}{q\pm \sqrt{\xi }}\right) 
\mathcal{G}^{(\pm )}(u),  \label{rs.17}
\end{equation}%
which allows write $k_{11}(u)$ as%
\begin{eqnarray}
k_{11}(u) &=&\frac{(1-\varepsilon _{2n}\theta _{2n}^{2}q^{2})\mathcal{F}%
_{n-1}^{(\pm )}(u)}{\mathrm{e}^{2u}-1}+\left\{ \beta _{n-1,n-1}-\beta
_{11}+(\beta _{2n,2n}-\beta _{n-1,n-1})\mathrm{e}^{u}\right.  \notag \\
&&\left. -(\beta _{2n,2n}-\beta _{11}-2)\mathrm{e}^{u}\left( \frac{q\mathrm{e%
}^{u}\pm \sqrt{\xi }}{q\pm \sqrt{\xi }}\right) \right\} \frac{\mathcal{G}%
^{(\pm )}(u)}{\mathrm{e}^{2u}-1}  \label{rs.18}
\end{eqnarray}

Substituting these expressions into the reflection equations we will get
constraint equations to fix some of the $3n-1$ remaining parameters. In
order to do this we recall the equations $E[2,2n+j]$ , ($j\neq n,n+1$) to
find $\beta _{jj}$, $j=3,4,...,2n-2$ in terms of $\beta _{22}$. Then the $%
\beta _{22}$ is given by the equations $E[2,2n+1]$. The parameters $\beta
_{n,n}$ and $\beta _{n+1,n+1}$ can be fixed in terms of $\beta _{22}$ using
the equations $E[n,2n^{2}-n+2]$ and $E[n+1,2n^{2}+n+2]$, respectively. After
this we can use the equation $E[2,2n-1]$ to find $\beta _{2n-1,2n-1}$ and $%
\beta _{2n,2n}$. These parameters can be written in terms of $\beta
_{11},\beta _{1,n-1},\beta _{1,n+2}$ and $\beta _{1,2n}$ in the following
way:%
\begin{equation}
\beta _{i,i}=\beta _{11}+\Delta _{n-1}^{(\pm
)}\sum_{j=0}^{i-2}(-q)^{j},\qquad 1<i\leq n-1  \label{rs.19}
\end{equation}%
\begin{equation}
\beta _{n,n}=\beta _{n+1,n+1}=\beta _{11}+\Delta _{n-1}^{(\pm )}\left[ \frac{%
1-(-q)^{n-2}}{q+1}\right] -q^{n-2}\Sigma _{n-1}^{(\pm )}  \label{rs.20}
\end{equation}%
\begin{equation}
\beta _{n+2,n+2}=\beta _{n+1,n+1}=\beta _{11}+\Delta _{n-1}^{(\pm )}\left[ 
\frac{1-(-q)^{n-2}}{q+1}\right] -(q^{n-2}-\varepsilon _{n+2}\theta
_{n+2}^{2}q^{n})\Sigma _{n-1}^{(\pm )}  \label{rs.21}
\end{equation}%
\begin{equation}
\beta _{ii}=\beta _{n+2,n+2}+\Delta _{n-1}^{(\pm )}\left[ \frac{\varepsilon
_{n+2}\theta _{n+2}^{2}}{\pm \sqrt{\xi }}\sum_{j=n-1}^{i-3}(-q)^{j}\ \right]
,\qquad n+3\leq i\leq 2n-1  \label{rs.22}
\end{equation}%
and%
\begin{equation}
\beta _{2n,2n}=\beta _{11}+2+\Delta _{n-1}^{(\pm )}\frac{(q\pm \sqrt{\xi })}{%
\xi }\frac{(\xi -\varepsilon _{2n}q^{\overset{\_}{2n}-\overset{\_}{1}})}{%
(q+1)^{2}}  \label{rs.23}
\end{equation}%
where%
\begin{equation}
\Sigma _{n-1}^{(\pm )}=\frac{\beta _{1,n-1}\beta _{1,n+2}}{\beta _{1,2n}}%
\frac{1}{\xi }\left( \frac{q\pm \sqrt{\xi }}{q+1}\right) ^{2}  \label{rs.24}
\end{equation}%
Next, we can use, for instance, the equation $E[2,4n]$ to fix $\beta
_{1,n-1} $: 
\begin{equation}
\beta _{1,n-1}=(-1)^{n-1}\frac{2\xi (\pm \sqrt{\xi })(q+1)^{2}}{(1\mp \sqrt{%
\xi })\left[ q^{n-1}\pm (-1)^{n}\sqrt{\xi }\right] \left[ \varepsilon
_{n+2}\theta _{n+2}^{2}q^{n}\pm (-1)^{n}\sqrt{\xi }\right] (q\pm \sqrt{\xi })%
}\frac{\beta _{1,2n}}{\beta _{1,n+2}}.  \label{rs.25}
\end{equation}%
Following the discussion made for the general solutions, we also need to
separate these solutions in order to take into account the existence of $%
k_{1,n-1}(u)$ for each model:

\begin{itemize}
\item For A$_{2n-1}^{(2)}$ models $\xi =-q^{2n}$ and $\theta
_{k}=\varepsilon _{k}=1,\forall k$. In this case there are no restrictions
in (\ref{rs.25}). It follows that the solution with $\mathcal{G}^{(+)}(u)$
(up sign) is related to the solution with $\mathcal{G}^{(-)}(u)$ (down sign)
by complex conjugation.

\item For D$_{n}^{(1)}$ models \ we have $\xi =q^{2n-2}$, $\theta
_{k}=\varepsilon _{k}=1,\forall k$. It means that the factors $\left[
q^{n-1}\pm (-1)^{n}\sqrt{\xi }\right] $ are different from zero for the
solution with $\mathcal{G}^{(+)}(u)$ only if $n$ is even and for the
solution with $\mathcal{G}^{(-)}(u)$ if $n$ is odd.

\item For C$_{n}^{(1)}$ models $\xi =q^{2n+2}$, $\varepsilon _{n+2}=-1$ and$%
\ \theta _{n+2}^{2}=q^{2}$. \ In this case there are also no restrictions
because both factors $\left[ q^{n-1}\pm (-1)^{n}\sqrt{\xi }\right] $ and $%
\left[ -q^{n+2}\pm (-1)^{n}\sqrt{\xi }\right] $ are different from zero. It
means that we have two independent solutions, one with $\mathcal{G}^{(+)}(u)$
and another with $\mathcal{G}^{(-)}(u)$, for all $n>3$.
\end{itemize}

Comparing these results with those flowed in the previous section, one could
conclude that we have simply made a reduction of the general solution
through an appropriate choice of the free parameters. Nevertheless, new
solutions are appearing for C$_{n}^{(1)}$ and D$_{n}^{(1)}$ models.

Substituting all fixed parameters into (\ref{rs.18}) we will find the
following expressions for the amplitude $k_{11}(u)$:

\begin{eqnarray}
k_{11}(u) &=&\frac{2\mathrm{e}^{u}\mathcal{G}^{(\pm )}(u)}{\mathrm{e}^{2u}-1}
\notag \\
&&-\frac{2\mathcal{G}^{(\pm )}(u)}{\mathrm{e}^{u}+1}\left\{ \frac{\xi \left(
1+q-(-q)^{n}+(-q)^{n-1}\right) +q\mathrm{e}^{u}(\xi -q^{2n-2})}{(1\mp \sqrt{%
\xi })\left[ q^{n-1}\pm (-1)^{n}\sqrt{\xi }\right] \left[ q^{n}\pm (-1)^{n}%
\sqrt{\xi }\right] }\right\}  \label{rs.26}
\end{eqnarray}%
for A$_{2n-1}^{(2)}$ and D$_{n}^{(1)}$ models and

\begin{eqnarray}
k_{11}(u) &=&\frac{2\mathrm{e}^{u}\mathcal{G}^{(\pm )}}{\mathrm{e}^{2u}-1} 
\notag \\
&&+\frac{2\mathcal{G}^{(\pm )}}{\mathrm{e}^{u}+1}\left\{ \frac{\xi \left(
1+q+(-q)^{n-1}+(-q)^{n+2}\right) +q\mathrm{e}^{u}\left( \xi +q^{2n}\right) }{%
(1\mp \sqrt{\xi })\left[ q^{n-1}\pm (-1)^{n}\sqrt{\xi }\right] \left[
q^{n+2}\mp (-1)^{n}\sqrt{\xi }\right] }\right\}  \label{rs.27}
\end{eqnarray}%
for C$_{n}^{(1)}$ models.

For D$_{n}^{(1)}$ case, we still have the simplified expression for $%
k_{11}(u)$ but now interchanging the parity of the solutions with $\mathcal{G%
}^{(\pm )}(u)$:%
\begin{equation}
k_{11}(u)=\frac{\mathcal{G}^{(+)}(u)}{\mathrm{e}^{u}-1}\quad (n-\mathrm{even}%
),\qquad k_{11}(u)=\frac{\mathcal{G}^{(-)}(u)}{\mathrm{e}^{u}-1}\quad (n-%
\mathrm{odd}).  \label{rs.28}
\end{equation}%
For C$_{n}^{(1)}$ models we loose the relations (\ref{gs.33}) between the
diagonal entries, but (\ref{rs.27}) defines a new solution with $\mathcal{G}%
^{(+)}(u)$ when $n$ is odd and another one with $\mathcal{G}^{(-)}(u)$ when $%
n$ is even.

Now, let us summarize these results: First, we have from (\ref{rs.1}) to (%
\ref{rs.10}) all non-diagonal matrix elements after substituting the $n-2$
fixed non-diagonal parameters $\beta _{21}$ and $\beta _{1,j}\ (j=2,...,n-2)$
given by (\ref{rs.11}) and (\ref{rs.25}).

Second, the diagonal matrix elements are obtained by using (\ref{rs.12}), (%
\ref{rs.13}) and (\ref{rs.14}) with $k_{11}(u)$ given by (\ref{rs.28}) for D$%
_{n}^{(1)}$ models and given by (\ref{rs.27}) for C$_{n}^{(1)}$ models and
by substituting the diagonal parameters given by (\ref{rs.19})-- (\ref{rs.23}%
). This sequence results in two solutions with $n$ parameters $\beta
_{1,n+2},\beta _{1,n+3},\ldots ,\beta _{1,2n}$ and $\beta _{11}$for\ these
models.

The number of free parameters in these new solutions is $n-1$ because we
still have to use the regular condition (\ref{re.10}), which will fix the
parameter $\beta _{11}$.

The general solutions which we have found in the previous section have $n$
free parameters. Therefore, solutions with $n-1$ free parameters can be
understood as reductions of them through a complicate limit procedure as was
described in this section. This is true for A$_{2n-1}^{(2)}$ models. For D$%
_{n}^{(1)}$ models our general solutions with $\mathcal{G}^{(+)}(u)$ have $n$
free parameters but only for $n=3,5,7,...$, and our general solutions with $%
\mathcal{G}^{(-)}(u)$ are defined only for $n=4,6,8,...\ $. The limit
procedure derived here gives us solutions with $\mathcal{G}^{(+)}(u)$ and $%
n-1$ free parameters provided that $n=4,6,8,...$ and solutions with $%
\mathcal{G}^{(-)}(u)$ and $n-1$ free parameters if $n=5,7,9,...$. Similar
considerations hold for C$_{n}^{(1)}$ models interchanging the parity of $n$.

Here we note that $n=3$ are solutions with $3$ free parameters and will be
treated in the section $6$.

We should continue with our reduction procedure in order to verify if exists
other solutions for those models. The next step is to consider $%
k_{1,n-1}(u)=0$, in addition to $k_{1,n}(u)=0$. In this case we will find
solutions with $16n-20$ null entries and with $n-2$ free parameters for all $%
n>4$. \ However, all these solutions are reductions of those solutions with $%
n$ and $n-1$ free parameters. So, in this step we have not found any new
solutions. Here the cases with $n=4$ will have $n-1$ free parameters.

The rational limit of some of these reductions was presented in \cite{AACDFR}%
.

After exhausting all the possible reductions we will arrive to the last
reduction which is obtained after applying $n-1$ reduction steps. The final
reduction for A$_{2n-1}^{(2)}$ and D$_{n}^{(1)}$ is the non-diagonal $K$%
-matrix solution of the reflection equation with the following entries
different from zero%
\begin{eqnarray}
k_{11}(u) &=&1,\qquad k_{2n,2n}(u)=\mathrm{e}^{2u},  \notag \\
k_{22}(u) &=&k_{33}(u)=\cdots k_{2n-1,2n-1}(u)=\frac{q^{2n-2}-\mathrm{e}^{2u}%
}{q^{2n-2}-1},  \notag \\
k_{1,2n}(u) &=&\frac{1}{2}\beta _{1,2n}\left( \mathrm{e}^{2u}-1\right)
,\qquad k_{2n,1}(u)=\frac{2}{\beta _{1,2n}}\frac{q^{2n-2}}{\left(
q^{2n-2}-1\right) ^{2}}\left( \mathrm{e}^{2u}-1\right) ,  \label{rs.29}
\end{eqnarray}%
while for C$_{n}^{(1)}$ the corresponding reduction is%
\begin{eqnarray}
k_{11}(u) &=&1,\qquad k_{2n,2n}(u)=\mathrm{e}^{2u},  \notag \\
k_{22}(u) &=&k_{33}(u)=\cdots k_{2n-1,2n-1}(u)=\frac{q^{2n}+\mathrm{e}^{2u}}{%
q^{2n}+1},  \notag \\
k_{1,2n}(u) &=&\frac{1}{2}\beta _{1,2n}\left( \mathrm{e}^{2u}-1\right)
,\qquad k_{2n,1}(u)=-\frac{2}{\beta _{1,2n}}\frac{q^{2n}}{\left(
q^{2n}+1\right) ^{2}}\left( \mathrm{e}^{2u}-1\right) .  \label{rs.30}
\end{eqnarray}%
Here we note that these final reductions are independent of $\mathcal{G}%
^{(\pm )}(u)$ and of $\xi $. Therefore they are new solutions.

In conclusion we have the following picture: for $n>2$

\begin{itemize}
\item For A$_{2n-1}^{(1)}$ models we have found three solutions: two $n$-
parameter conjugated solutions with all entries different from zero and one $%
1$-parameter "real" solution given by (\ref{rs.29}).

\item For D$_{n}^{(1)}$ models\ we have found three solutions: one $n$%
-parameter solution with all entries different from zero, one ($n-1$%
)-parameter solution with $8n-6$ null entries both $n$-parity dependent, and
one $1$-parameter solution given by (\ref{rs.29}).

\item For C$_{n}^{(1)}$ models\ we have found three solutions: one $n$%
-parameter solution with all entries different from zero, one ($n-1$%
)-parameter solution with $8n-6$ null entries both $n$-parity dependent, and
one $1$-parameter solution given by (\ref{rs.30}).
\end{itemize}

These results give us all independent solutions of the reflection equations (%
\ref{int.2}), with at least two non-diagonal entries. Nevertheless, special
attention is reserved for reductions which give us diagonal $K$-matrix
solutions. \ 

\section{Diagonal K-Matrix Solutions}

The main goal here is to present all diagonal solutions from the reflection
equations (\ref{int.2}) and (\ref{int.3}) for $\mathcal{R}$-matrices
associated to C$_{n}^{(1)}$, D$_{n}^{(1)}$ and A$_{2n-1}$ affine Lie
algebras. Instead of using a reduction procedure, we choose to find the
diagonal solutions by solving again the reflection equations. In this
section we will consider these solutions for each model separately.

\subsection{A$_{2n-1}^{(2)}$ Diagonal K-matrices}

Solving the reflection equations we find one trivial solution which is
proportional to the identity, two $1$-parameter solution $\mathbb{K}_{\beta
} $ with the following normalized matrix elements: 
\begin{eqnarray}
k_{11}(u) &=&k_{22}(u)=\cdots =k_{n-1,n-1}(u)=1,  \notag \\
k_{n,n}(u) &=&\left( \frac{\beta (\mathrm{e}^{u}-1)-2}{\beta (\mathrm{e}%
^{-u}-1)-2}\right) ,  \notag \\
k_{n+1,n+1}(u) &=&\mathrm{e}^{2u}\left( \frac{\beta (\mathrm{e}%
^{-u}+q^{2n-2})+2q^{2n-2}}{\beta (\mathrm{e}^{u}+q^{2n-2})+2q^{2n-2}}\right)
,  \notag \\
k_{2n,2n}(u) &=&k_{2n-1,2n-1}(u)=\cdots =k_{n+2,n+2}(u)=\mathrm{e}^{2u},
\label{ds.1}
\end{eqnarray}%
where $\beta $ is the free parameter. The second solution is obtained from (%
\ref{ds.1}) using the symmetry $k_{n,n}(u)$ $\longleftrightarrow $ $%
k_{n+1,n+1}(u)$. Moreover, for $n>2$ we find $2n-4$ solutions\ $\mathbb{K}%
^{[p]}$\ , $p=2,3,\ldots ,n-1$ without any free parameters and with the
following entries:%
\begin{eqnarray}
k_{11}(u) &=&k_{22}(u)=\cdots =k_{p-1,p-1}(u)=1,  \notag \\
k_{p,p}(u) &=&k_{p+1,p+1}(u)=\cdots =k_{2n-p+1,2n-p+1}(u)=\mathrm{e}^{2u}%
\frac{\mathrm{e}^{-u}\pm iq^{2p-n-1}}{\mathrm{e}^{u}\pm q^{2p-n-1}},  \notag
\\
k_{2n,2n}(u) &=&k_{2n-1,2n-1}(u)=\cdots =k_{2n-p+2,2n-p+2}(u)=\mathrm{e}%
^{2u},  \notag \\
p &=&2,3,\ldots ,n-1.  \label{ds.2}
\end{eqnarray}

Here we observe that: the cases $p=n$ are not computed as solutions because
the $\mathbb{K}^{[p=n]}$ solutions are obtained from $\mathbb{K}_{\beta }$
by the choice of $\beta $ such that $k_{n,n}=k_{n+1,n+1}$. Therefore, for
the A$_{2n-1}^{(2)}$ models we have found $2n-1$ regular diagonal solutions.

We mention that the case $n=2$ has been first discussed by Martins in the
context of coupled six vertex models \cite{MA}.

\subsection{D$_{n}^{(1)}$ Diagonal K-matrices}

For these models we also have the symmetry $k_{n,n}(u)\longleftrightarrow
k_{n+1,n+1}(u)$ in the $K$-matrix solutions.

The D$_{2}^{(1)}$ diagonal solutions are special. In addition to the
identity we have found two solutions with two free parameters%
\begin{equation}
\mathbb{K}_{\alpha \beta }=\left( 
\begin{array}{cccc}
1 & 0 & 0 & 0 \\ 
0 & \frac{\alpha \left( \mathrm{e}^{u}-1\right) -2}{\alpha \left( \mathrm{e}%
^{-u}-1\right) -2} & 0 & 0 \\ 
0 & 0 & \frac{\beta \left( \mathrm{e}^{u}-1\right) -2}{\beta \left( \mathrm{e%
}^{-u}-1\right) -2} & 0 \\ 
0 & 0 & 0 & \frac{\alpha \left( \mathrm{e}^{u}-1\right) -2}{\alpha \left( 
\mathrm{e}^{-u}-1\right) -2}\frac{\beta \left( \mathrm{e}^{u}-1\right) -2}{%
\beta \left( \mathrm{e}^{-u}-1\right) -2}%
\end{array}%
\right)   \label{ds.3}
\end{equation}%
where $\alpha $ and $\beta $ are the free parameters. \ This $so(4)$
solution, in the isotropic limit, was already presented in \cite{AACDFR}.

For $n>2$ we have found the identity and seven $1$-parameter solutions $%
\mathbb{K}_{\beta }^{[i]}$ ($i=1,2,....,7$):

\begin{itemize}
\item The $\mathbb{K}_{\beta }^{[1]}$ matrix has the following entries:%
\begin{eqnarray}
k_{11}(u) &=&1,  \notag \\
k_{22}(u) &=&k_{33}(u)=\cdots =k_{2n-1,2n-1}(u)=\frac{\beta \left( \mathrm{e}%
^{u}-1\right) -2}{\beta \left( \mathrm{e}^{-u}-1\right) -2},  \notag \\
k_{2n,2n}(u) &=&\frac{\beta \left( \mathrm{e}^{u}-1\right) -2}{\beta \left( 
\mathrm{e}^{-u}-1\right) -2}\frac{\beta \left( q^{2n-4}\mathrm{e}%
^{u}-1\right) -2}{\beta \left( q^{2n-4}\mathrm{e}^{-u}-1\right) -2}.
\label{ds.4}
\end{eqnarray}

\item The $\mathbb{K}_{\beta }^{[2]}$ matrix has the following entries:%
\begin{eqnarray}
k_{11}(u) &=&k_{22}(u)=\cdots =k_{n,n}(u)=1  \notag \\
k_{n+1,n+1}(u) &=&k_{n+2,n+2}(u)=\cdots =k_{2n,2n}(u)=\frac{\beta \left( 
\mathrm{e}^{u}-1\right) -2}{\beta \left( \mathrm{e}^{-u}-1\right) -2}
\label{ds.5}
\end{eqnarray}

\item The $\mathbb{K}_{\beta }^{[3]}$ matrix has the following entries:%
\begin{eqnarray}
k_{11}(u) &=&k_{22}(u)=\cdots =k_{n-1,n-1}(u)=k_{n+1,n+1}(u)=1,  \notag \\
k_{n,n}(u) &=&k_{n+2,n+2}(u)=\cdots =k_{2n,2n}(u)=\frac{\beta \left( \mathrm{%
e}^{u}-1\right) -2}{\beta \left( \mathrm{e}^{-u}-1\right) -2}.  \label{ds.6}
\end{eqnarray}

\item The $\mathbb{K}_{\beta }^{[4]}$ matrix has the following entries:%
\begin{eqnarray}
k_{11}(u) &=&1,  \notag \\
k_{22}(u) &=&k_{33}(u)=\cdots =k_{n,n}(u)=\frac{\beta \left( \mathrm{e}%
^{u}-1\right) -2}{\beta \left( \mathrm{e}^{-u}-1\right) -2},  \notag \\
k_{n+1,n+1}(u) &=&k_{n+2,n+2}(u)=\cdots =k_{2n-1,2n-1}(u)=\mathrm{e}^{2u}, 
\notag \\
k_{2n,2n}(u) &=&\mathrm{e}^{2u}\frac{\beta \left( \mathrm{e}^{u}-1\right) -2%
}{\beta \left( \mathrm{e}^{-u}-1\right) -2}.  \label{ds.7}
\end{eqnarray}

\item The $\mathbb{K}_{\beta }^{[5]}$ matrix has the following entries:%
\begin{eqnarray}
k_{11}(u) &=&1,  \notag \\
k_{22}(u) &=&k_{33}(u)=\cdots =k_{n-1,n-1}(u)=k_{n+1,n+1}(u)=\frac{\beta
\left( \mathrm{e}^{u}-1\right) -2}{\beta \left( \mathrm{e}^{-u}-1\right) -2},
\notag \\
k_{n,n}(u) &=&k_{n+2,n+2}(u)=\cdots =k_{2n-1,2n-1}(u)=\mathrm{e}^{2u}, 
\notag \\
k_{2n,2n}(u) &=&\mathrm{e}^{2u}\frac{\beta \left( \mathrm{e}^{u}-1\right) -2%
}{\beta \left( \mathrm{e}^{-u}-1\right) -2}.  \label{ds.8}
\end{eqnarray}

\item The $\mathbb{K}_{\beta }^{[6]}$ matrix has the following entries:%
\begin{eqnarray}
k_{11}(u) &=&k_{22}(u)=\cdots =k_{n-1,n-1}(u)=1,  \notag \\
k_{n,n}(u) &=&\mathrm{e}^{2u}\frac{\beta \left( \mathrm{e}%
^{-u}-q^{2n-4}\right) -2q^{2n-4}}{\beta \left( \mathrm{e}^{u}-q^{2n-4}%
\right) -2q^{2n-4}},  \notag \\
k_{n+1,n+1}(u) &=&\frac{\beta \left( \mathrm{e}^{u}-1\right) -2}{\beta
\left( \mathrm{e}^{-u}-1\right) -2}.  \notag \\
k_{n+2,n+2}(u) &=&k_{n+3,n+3}(u)=\cdots =k_{2n,2n}(u)=\mathrm{e}^{2u}.
\label{ds.9}
\end{eqnarray}

\item The $\mathbb{K}_{\beta }^{[7]}$ matrix has the following entries:%
\begin{eqnarray}
k_{11}(u) &=&k_{22}(u)=\cdots =k_{n-1,n-1}(u)=1,  \notag \\
k_{n,n}(u) &=&\frac{\beta \left( \mathrm{e}^{u}-1\right) -2}{\beta \left( 
\mathrm{e}^{-u}-1\right) -2},  \notag \\
k_{n+1,n+1}(u) &=&\mathrm{e}^{2u}\frac{\beta \left( \mathrm{e}%
^{-u}-q^{2n-4}\right) -2q^{2n-4}}{\beta \left( \mathrm{e}^{u}-q^{2n-4}%
\right) -2q^{2n-4}},  \notag \\
k_{n+2,n+2}(u) &=&k_{n+3,n+3}(u)=\cdots =k_{2n,2n}(u)=\mathrm{e}^{2u}.
\label{ds.10}
\end{eqnarray}
\end{itemize}

Moreover, for $n>3$ we have found $n-3$ solutions without any free
parameters $\mathbb{K}^{[p]}$, $p=3,4,...,n-1$, with the following matrix
elements%
\begin{eqnarray}
k_{11}(u) &=&k_{22}(u)=\cdots =k_{p-1,p-1}(u)=1,  \notag \\
k_{p,p}(u) &=&k_{p+1,p+1}(u)=\cdots =k_{2n-p+1,2n-p+1}(u)=\mathrm{e}^{2u}%
\frac{\mathrm{e}^{-u}+\epsilon _{p}q^{2p-n-2}}{\mathrm{e}^{u}+\epsilon
_{p}q^{2p-n-2}},  \notag \\
k_{2n-p+2,2n-p+2}(u) &=&k_{2n-p+3,2n-p+3}(u)=\cdots =k_{2n,2n}(u)=\mathrm{e}%
^{2u},  \label{ds.11}
\end{eqnarray}%
where $\epsilon _{p}=\pm 1$ for $2p\neq n+2$ and $\epsilon _{p}=1$ for $%
2p=n+2$. Therefore, for $n\geq 3$ the D$_{n}^{(1)}$ models have $2n+1$
diagonal solutions if $n$ is odd and $2n$ diagonal solutions if $n$ is even.

Here we note that the cases $\mathbb{K}^{[p=n]}$ are not computed here
because they are reductions of the cases $\mathbb{K}_{\beta }^{[6]}$ and $%
\mathbb{K}_{\beta }^{[7]}$ by the choice of $\beta $ such that $%
k_{n,n}(u)=k_{n+1,n+1}(u)$. Moreover, the cases $\mathbb{K}^{[p=2]}$ are
reductions of $\mathbb{K}_{\beta }^{[1]}$ with the choice of $\beta $ such
that $k_{2n,2n}(u)=\mathrm{e}^{2u}$.

\subsection{C$_{n}^{(1)}$ Diagonal K-matrices}

The $K$-matrix solutions for these models do not enjoy the symmetry $%
k_{n,n}(u)\longleftrightarrow k_{n+1,n+1}(u)$. After solving the
corresponding reflection equations we have found one solution with one free
parameter $\mathbb{K}_{\beta }$, with the following entries:%
\begin{eqnarray}
k_{11}(u) &=&k_{22}(u)=\cdots =k_{n,n}(u)=1,  \notag \\
k_{n+1,n+1}(u) &=&k_{n+2,n+2}(u)=\cdots =k_{2n,2n}(u)=\frac{\beta \left( 
\mathrm{e}^{u}-1\right) -2}{\beta \left( \mathrm{e}^{-u}-1\right) -2},
\label{ds.12}
\end{eqnarray}%
where $\beta $ is the free parameter. \ For $n>2$, in addition to the
identity we also have found $n-1$ solutions without \ any free parameters $%
\mathbb{K}^{[p]}$, $p=2,3,...,n$, with the following matrix elements:%
\begin{eqnarray}
k_{11}(u) &=&k_{22}(u)=\cdots =k_{p-1,p-1}(u)=1,  \notag \\
k_{p,p}(u) &=&k_{p+1,p+1}(u)=\cdots =k_{2n-p+1,2n-p+1}(u)=\mathrm{e}^{2u}%
\frac{\mathrm{e}^{-u}+\epsilon _{p}q^{2p-n-2}}{\mathrm{e}^{u}+\epsilon
_{p}q^{2p-n-2}},  \notag \\
k_{2n-p+2,2n-p+2}(u) &=&k_{2n-p+3,2n-p+3}(u)=\cdots =k_{2n,2n}(u)=\mathrm{e}%
^{2u},  \label{ds.13}
\end{eqnarray}%
where $\epsilon _{p}=\pm 1$ for $2p\neq n+2$ and $\epsilon _{p}=1$ for $%
2p=n+2$. Therefore, for the C$_{n}^{(1)}$ models we have $2n$ diagonal
solutions if $n$ is odd and $2n-1$ diagonal solutions if $n$ is even.

Here we would like to note that our C$_{n}^{(1)}$ diagonal solutions are
contained in the set of the D$_{n}^{(1)}$ diagonal solutions.

A final note about these diagonal solutions is that almost all those with
free parameters degenerate, after an appropriate choice of the free
parameters, into the two type of solutions presented in \textit{\cite{Batch2}%
}. Namely the identity ($s=1$) and the $K$ matrix ($s=0$) with the following
normalized entries:%
\begin{eqnarray}
k_{11}(u) &=&k_{22}(u)=\cdots =k_{n,n}(u)=1  \label{ds.14} \\
k_{n+1,n+1}(u) &=&k_{n+2,n+2}(u)=\cdots =k_{2n,2n}(u)=\mathrm{e}^{2u}  \notag
\end{eqnarray}

\section{Special Cases}

In this section we will present solutions which are out of our
classification scheme. They are the case $n=1$, the case $n=2$ , the
solution with $\mathcal{G}^{(-)}(u)$ for D$_{3}^{(1)}$ and the solution with 
$\mathcal{G}^{(+)}(u)$ for C$_{3}^{(1)}$:

\subsection{C$_{1}^{(1)}$, D$_{1}^{(1)}$ and A$_{1}^{(2)}$ K-matrix solutions%
}

These models have one common general solution with three free parameters $%
\beta _{11},\beta _{12}$ and $\beta _{21}$

\begin{equation}
K^{-}=\left( 
\begin{array}{cc}
1+\beta _{11}(\mathrm{e}^{u}-1) & \frac{1}{2}\beta _{12}(\mathrm{e}^{2u}-1)
\\ 
\frac{1}{2}\beta _{21}(\mathrm{e}^{2u}-1) & \mathrm{e}^{2u}-\beta _{11}%
\mathrm{e}^{u}(\mathrm{e}^{u}-1)%
\end{array}%
\right)  \label{sp.0}
\end{equation}%
and one common diagonal solution which can be obtained from (\ref{sp.0})
when $\beta _{12}=\beta _{21}=0$ and the identity.

\subsection{ A$_{3}^{(2)}$ K-matrix solutions}

In this model we have found one general solution with four free parameters, $%
\beta _{12},\beta _{13},\beta _{14}$ and $\beta _{24}$. The $K^{-}$ matrix
has the form

\begin{equation}
K^{-}=\left( 
\begin{array}{cccc}
k_{11} & k_{12} & k_{13} & k_{14} \\ 
k_{21} & k_{22} & k_{23} & k_{24} \\ 
k_{31} & k_{32} & k_{33} & k_{34} \\ 
k_{41} & k_{42} & k_{43} & k_{44}%
\end{array}%
\right)  \label{sp.1}
\end{equation}%
where the normalized diagonal entries are given by%
\begin{eqnarray*}
k_{22} &=&\mathrm{e}^{u}+\frac{\beta _{12}\mathrm{e}^{u}(\mathrm{e}^{u}-1)}{%
2q^{2}\beta _{14}\beta _{24}(q^{2}+\mathrm{e}^{2u})}\left\{ \beta _{24}\left[
q^{2}\left( \beta _{13}+\beta _{24}\right) -\beta _{13}\right] \left( 
\mathrm{e}^{u}+q^{2}\right) \right. \\
&&\left. +\beta _{13}^{2}\left( q^{2}\mathrm{e}^{u}+1\right) \right\}
\end{eqnarray*}%
\begin{eqnarray*}
k_{33} &=&\mathrm{e}^{u}-\frac{\beta _{12}\mathrm{e}^{u}(\mathrm{e}^{u}-1)}{%
2q^{2}\beta _{14}\beta _{24}(q^{2}+\mathrm{e}^{2u})}\left\{ \beta _{13}\left[
\left( \beta _{13}+\beta _{24}\right) -q^{2}\beta _{24}\right] \left( 
\mathrm{e}^{u}+q^{2}\right) \right. \\
&&\left. +\beta _{24}^{2}q^{2}\left( q^{2}\mathrm{e}^{u}+1\right) \right\}
\end{eqnarray*}%
\begin{eqnarray}
k_{44} &=&\mathrm{e}^{u}+\frac{\beta _{12}\mathrm{e}^{u}(\mathrm{e}^{u}-1)}{%
2q^{2}\beta _{14}\beta _{24}(q^{2}+\mathrm{e}^{2u})}\left\{ \beta _{13}\beta
_{24}\left[ \left( q^{2}+\mathrm{e}^{u}\right) ^{2}-q^{2}\left( \mathrm{e}%
^{2u}-1\right) \right] \right.  \notag \\
&&\left. +\left( \beta _{13}^{2}-q^{2}\beta _{24}^{2}\right) \mathrm{e}%
^{u}\left( q^{2}+\mathrm{e}^{u}\right) \right\}  \label{sp.2}
\end{eqnarray}%
and the non-diagonal matrix elements are

\begin{eqnarray}
k_{12} &=&\frac{\beta _{12}}{2\beta _{24}}f(u),\quad k_{13}=\frac{1}{2}%
g(u),\quad k_{14}=\frac{1}{2}\beta _{14}\left( \mathrm{e}^{2u}-1\right) , 
\notag \\
\quad k_{21} &=&-\frac{1}{2}\ \Omega f(u),\quad k_{23}=\frac{\beta
_{14}\beta _{24}}{2\beta _{12}}\ \Omega \left( \mathrm{e}^{2u}-1\right)
,\quad k_{24}=\frac{1}{2}\mathrm{e}^{u}f(u),  \notag \\
k_{31} &=&\frac{\beta _{12}}{2q^{2}\beta _{24}}\ \Omega g(u),\quad k_{32}=-%
\frac{\beta _{12}\beta _{14}}{2q^{2}\beta _{24}}\ \Omega \left( \mathrm{e}%
^{2u}-1\right) ,\quad k_{34}=-\frac{\beta _{12}}{2q^{2}\beta _{24}}\mathrm{e}%
^{u}g(u),  \notag \\
\quad k_{41} &=&-\frac{\beta _{14}}{2q^{2}}\ \Omega ^{2}\left( \mathrm{e}%
^{2u}-1\right) ,\quad k_{42}=\frac{\beta _{12}}{2q^{2}\beta _{24}}\ \Omega 
\mathrm{e}^{u}f(u),\quad k_{43}=\frac{1}{2q^{2}}\ \Omega \mathrm{e}^{u}g(u).
\label{sp.3}
\end{eqnarray}%
Here we have defined two scalar functions $f(u)$ and $g(u)$ different from
our $\mathcal{G}^{(\pm )}(u)$ functions: 
\begin{eqnarray}
f(u) &=&\left[ \beta _{13}(\mathrm{e}^{u}-1)+\beta _{24}(\mathrm{e}%
^{u}+q^{2})\right] \left( \frac{\mathrm{e}^{2u}-1}{\mathrm{e}^{2u}+q^{2}}%
\right) ,  \notag \\
g(u) &=&\left[ \beta _{13}(\mathrm{e}^{u}+q^{2})-q^{2}\beta _{24}(\mathrm{e}%
^{u}-1)\right] \left( \frac{\mathrm{e}^{2u}-1}{\mathrm{e}^{2u}+q^{2}}\right)
,  \notag \\
\Omega &=&\frac{\beta _{12}\beta _{13}(q^{2}+1)-2q^{2}\beta _{14}}{\beta
_{14}^{2}(q^{2}+1)}.  \label{sp.4}
\end{eqnarray}%
This solution can be unfolded in the two parameter general solutions with $%
\mathcal{G}^{(\pm )}(u)$, by an appropriate choice of $\beta _{12}$ and $%
\beta _{24}$. It follows, by the reduction procedure, the solution (\ref%
{rs.29}) and the two non-trivial diagonal matrices given by (\ref{ds.1}).

\subsection{C$_{2}^{(1)}$ K-matrix Solutions}

For this case we have one general solution with three free parameters. The $%
K^{-}$ matrix has the form

\begin{equation}
K^{-}=\left( 
\begin{array}{cccc}
k_{11} & k_{12} & k_{13} & k_{14} \\ 
k_{21} & k_{22} & k_{23} & \frac{\mathrm{e}^{u}}{q^{2}}k_{13} \\ 
k_{31} & k_{32} & \mathrm{e}^{u}k_{11} & -\mathrm{e}^{u}k_{12} \\ 
k_{41} & \frac{\mathrm{e}^{u}}{q^{2}}k_{31} & -\mathrm{e}^{u}k_{21} & 
\mathrm{e}^{u}k_{22}%
\end{array}%
\right) ,  \label{sp.5}
\end{equation}%
with the following remaining diagonal entries: 
\begin{equation}
k_{11}=1-\frac{\beta _{12}\beta _{13}}{q^{2}\beta _{14}}f(u),\quad k_{22}=1+%
\frac{\beta _{12}\beta _{13}}{\beta _{14}}f(u).  \label{sp.6}
\end{equation}%
\begin{eqnarray}
k_{12} &=&\beta _{12}f(u),\quad k_{13}=\beta _{13}f(u),\quad k_{14}=\beta
_{14}(\mathrm{e}^{u}-1),  \notag \\
k_{21} &=&-\frac{\beta _{13}}{q^{2}\beta _{14}}\ \Gamma f(u),\quad k_{23}=%
\frac{\beta _{13}}{q^{2}\beta _{12}}\ \Gamma (\mathrm{e}^{u}-1),  \notag \\
k_{31} &=&\frac{\beta _{12}}{\beta _{14}}\ \Gamma f(u),\quad k_{32}=-\frac{%
\beta _{12}}{\beta _{13}}\ \Gamma (\mathrm{e}^{u}-1),\quad k_{41}=-\frac{%
\Gamma ^{2}}{q^{2}\beta _{14}}(\mathrm{e}^{u}-1).  \label{sp.7}
\end{eqnarray}%
Here we also have used a new scalar function $f(u)$, a little bit different
from $\mathcal{G}^{(+)}(u)$: 
\begin{equation}
f(u)=\left( \frac{q^{2}+1}{q^{2}+\mathrm{e}^{u}}\right) (\mathrm{e}%
^{u}-1)\qquad \mathrm{and\qquad }\Gamma =\frac{\beta _{12}\beta _{13}}{\beta
_{14}}-\frac{q^{2}}{q^{2}+1}.  \label{sp.8}
\end{equation}%
This solution can be identified with the two parameter general solution with 
$\mathcal{G}^{(+)}(u)$ by an appropriate choice of $\beta _{12}$. It
follows, by the reduction procedure, the solution (\ref{rs.30}) and the
three diagonal matrices given by (\ref{ds.12}) and (\ref{ds.13}).

\subsection{D$_{2}^{(1)}$ K-matrix solutions}

This is a very special case because we do not have any solution with all
entries different from zero. Here we have the identity and one $2$-parameter
diagonal solution which is given by (\ref{ds.3}). In addition to the reduced
solution (\ref{rs.29}) we also have found the following $K^{-}$ matrix

\begin{equation}
K^{-}=\left( 
\begin{array}{cccc}
\mathrm{e}^{-u}\frac{q^{2}-\mathrm{e}^{2u}}{q^{2}-1} & 0 & 0 & 0 \\ 
0 & \mathrm{e}^{u} & \frac{1}{2}\beta \left( \mathrm{e}^{2u}-1\right) & 0 \\ 
0 & \frac{2q^{2}(\mathrm{e}^{2u}-1)}{\beta (q^{2}-1)^{2}} & \mathrm{e}^{u} & 
0 \\ 
0 & 0 & 0 & \mathrm{e}^{u}\frac{q^{2}-\mathrm{e}^{2u}}{q^{2}-1}%
\end{array}%
\right)  \label{sp.9}
\end{equation}%
where $\beta $ is a free parameter.

In this way we have listed all solutions of (\ref{int.2}) for C$_{2}^{(1)}$,
D$_{2}^{(1)}$ and A$_{3}^{(2)}$ models.\ The corresponding $K^{+}$ solutions
are obtained using (\ref{int.4}) and taking into account the $M$ matrix (\ref%
{re.15}) for each model.

\subsection{The D$_{3}^{(1)}$ solution with $\mathcal{G}^{(-)}(u)$}

In section $4$ we have found one $3$-parameter general solution with $%
\mathcal{G}^{(+)}(u)$ for this model. The reduction procedure gives us
another solution with $\mathcal{G}^{(-)}(u)$ which also has three free
parameters. The corresponding $K^{-}$ matrix has the form

\begin{equation}
K^{-}=\left( 
\begin{array}{cccccc}
k_{11} & k_{12} & 0 & 0 & k_{15} & k_{16} \\ 
k_{21} & k_{22} & 0 & 0 & k_{25} & k_{26} \\ 
0 & 0 & k_{33} & 0 & 0 & 0 \\ 
0 & 0 & 0 & k_{44} & 0 & 0 \\ 
k_{51} & k_{52} & 0 & 0 & k_{55} & k_{56} \\ 
k_{61} & k_{62} & 0 & 0 & k_{65} & k_{66}%
\end{array}%
\right)  \label{sp.10}
\end{equation}%
with the non-normalized diagonal matrix elements given by%
\begin{eqnarray}
k_{11} &=&\left( \frac{2(\mathrm{e}^{u}-q)}{(1-q)(\mathrm{e}^{u}-1)}+\frac{%
(1+q^{2})\beta _{12}\beta _{15}}{q\beta _{16}}\right) \frac{\mathcal{G}%
^{(-)}(u)}{\mathrm{e}^{u}+1},  \notag \\
k_{22} &=&\left( \frac{2(\mathrm{e}^{u}-q)}{(1-q)(\mathrm{e}^{u}-1)}+\frac{%
\left[ \mathrm{e}^{u}(q-1)+q(1+q)\right] \beta _{12}\beta _{15}}{q\beta _{16}%
}\right) \frac{\mathcal{G}^{(-)}(u)}{\mathrm{e}^{u}+1}.  \notag \\
k_{33} &=&k_{44}=\left( \frac{2(\mathrm{e}^{u}-q)^{2}}{(1-q^{2})(\mathrm{e}%
^{u}-1)}+\frac{(\mathrm{e}^{u}-q^{2})\beta _{12}\beta _{15}}{q\beta _{16}}%
\right) \frac{\mathrm{e}^{u}+q}{1-q}\frac{\mathcal{G}^{(-)}(u)}{\mathrm{e}%
^{u}+1},  \notag \\
k_{55} &=&\left( \frac{2(\mathrm{e}^{u}-q)\mathrm{e}^{u}}{(1-q)(\mathrm{e}%
^{u}-1)}+\frac{\left[ \mathrm{e}^{u}(q+1)+q(1-q)\right] \beta _{12}\beta
_{15}}{q\beta _{16}}\right) \frac{\mathrm{e}^{u}\mathcal{G}^{(-)}(u)}{%
\mathrm{e}^{u}+1},  \notag \\
k_{66} &=&\mathrm{e}^{2u}k_{11}.  \label{sp.11}
\end{eqnarray}%
and the non-diagonal entries are%
\begin{eqnarray}
k_{12} &=&\beta _{12}\mathcal{G}^{(-)}(u),\quad k_{15}=\beta _{15}\mathcal{G}%
^{(-)}(u),\quad k_{16}=\beta _{16}\frac{\mathrm{e}^{u}-q}{1-q}\mathcal{G}%
^{(-)}(u),  \notag \\
k_{21} &=&\beta _{21}\mathcal{G}^{(-)}(u),\quad k_{25}=-\beta _{21}\frac{%
\beta _{16}}{\beta _{12}}\frac{\mathrm{e}^{u}-q}{1-q}\mathcal{G}%
^{(-)}(u),\quad k_{26}=-\frac{\beta _{15}}{q}\mathrm{e}^{u}\mathcal{G}%
^{(-)}(u),  \notag \\
k_{51} &=&\beta _{21}\frac{q^{2}\beta _{12}}{\beta _{15}}\mathcal{G}%
^{(-)}(u),\quad k_{52}=-\beta _{21}\frac{q^{2}\beta _{12}\beta _{16}}{\beta
_{15}^{2}}\frac{\mathrm{e}^{u}-q}{1-q}\mathcal{G}^{(-)}(u),\quad
k_{56}=-q\beta _{12}\mathrm{e}^{u}\mathcal{G}^{(-)}(u),  \notag \\
k_{61} &=&\beta _{21}^{2}\frac{q^{2}\beta _{16}}{\beta _{15}^{2}}\frac{%
\mathrm{e}^{u}-q}{1-q}\mathcal{G}^{(-)}(u),\quad k_{62}=-\beta _{21}\frac{%
q\beta _{12}}{\beta _{15}}\mathrm{e}^{u}\mathcal{G}^{(-)}(u),\quad
k_{65}=-q\beta _{21}\mathrm{e}^{u}\mathcal{G}^{(-)}(u).  \notag \\
&&  \label{sp.12}
\end{eqnarray}%
where%
\begin{equation}
\mathcal{G}^{(-)}(u)=\frac{1}{\beta _{16}}\left( \frac{1-q}{e^{u}-q}\right)
k_{16}\qquad \mathrm{and\qquad }\beta _{21}=\frac{\beta _{12}\beta _{15}^{2}%
}{q\beta _{16}^{2}}-\frac{2}{q^{2}-1}\frac{\beta _{15}}{\beta _{16}}
\label{sp.13}
\end{equation}%
Here we note that for $n>3$, this type of solution has $n-1$ free parameters
and follows the classification scheme presented in section $4$.

\subsection{The C$_{3}^{(1)}$ solution with $\mathcal{G}^{(+)}(u)$}

For this model the general solution is given in terms of $\mathcal{G}%
^{(-)}(u)$ and also has three free parameters. Here we have another $3$%
-parameter solution in terms of $\mathcal{G}^{(+)}(u)$. It has the form (\ref%
{sp.10}) with the following non-normalized diagonal entries%
\begin{eqnarray}
k_{11} &=&\frac{2\mathrm{e}^{u}\mathcal{G}^{(+)}(u)}{\mathrm{e}^{2u}-1}%
-\left( \frac{\beta _{12}\beta _{15}}{\beta _{16}}\frac{1+q^{3}}{q^{3}}-%
\frac{q\beta _{21}\beta _{16}\left[ (1+q^{2})\mathrm{e}^{u}-q^{2}(1+q^{4})%
\right] }{\beta _{15}(1+q^{3})}\right) \frac{\mathcal{G}^{(+)}(u)}{\mathrm{e}%
^{u}+1},  \notag \\
k_{22} &=&\frac{2\mathrm{e}^{u}\mathcal{G}^{(+)}(u)}{\mathrm{e}^{2u}-1}%
+\left( \frac{\beta _{12}\beta _{15}}{\beta _{16}}\frac{(1+q^{3})\mathrm{e}%
^{u}}{q^{3}}+\frac{q\beta _{21}\beta _{16}\left[ (1+q^{2})\mathrm{e}%
^{u}-q^{2}(1+q^{4})\right] }{\beta _{15}(1+q^{3})}\right) \frac{\mathcal{G}%
^{(+)}(u)}{\mathrm{e}^{u}+1},  \notag \\
k_{33} &=&k_{44}=\frac{2\mathrm{e}^{u}\mathcal{G}^{(+)}(u)}{\mathrm{e}^{2u}-1%
}  \notag \\
&&+\left( \frac{\beta _{12}\beta _{15}}{\beta _{16}}\frac{(1+q^{3})\mathrm{e}%
^{u}}{q^{3}}+\frac{\beta _{21}\beta _{16}\left[ (1+q)(\mathrm{e}^{u}-q^{6})+(%
\mathrm{e}^{u}+q^{3})^{2}\right] }{\beta _{15}(1+q^{3})}\right) \frac{%
\mathcal{G}^{(+)}(u)}{\mathrm{e}^{u}+1},  \notag \\
k_{55} &=&\frac{2\mathrm{e}^{u}\mathcal{G}^{(+)}(u)}{\mathrm{e}^{2u}-1} 
\notag \\
&&+\left( \frac{\beta _{12}\beta _{15}}{\beta _{16}}\frac{1+q^{3}}{q^{3}}+%
\frac{q\beta _{21}\beta _{16}\left[ (1+q^{4})(\mathrm{e}%
^{u}+q^{3})+(1+q)(1+q^{3})\right] }{\beta _{15}(1+q^{3})}\right) \frac{%
\mathrm{e}^{u}\mathcal{G}^{(+)}(u)}{\mathrm{e}^{u}+1},  \notag \\
k_{66} &=&\frac{2\mathrm{e}^{3u}\mathcal{G}^{(+)}(u)}{\mathrm{e}^{2u}-1} 
\notag \\
&&-\left( \frac{\beta _{12}\beta _{15}}{\beta _{16}}\frac{(1+q^{3})\mathrm{e}%
^{u}}{q^{3}}+\frac{q\beta _{21}\beta _{16}\left[ (1+q^{2})(\mathrm{e}%
^{u}+q^{3})+q^{2}(1+q)(1+q^{3})\mathrm{e}^{u}\right] }{\beta _{15}(1+q^{3})}%
\right) \frac{\mathrm{e}^{u}\mathcal{G}^{(+)}(u)}{\mathrm{e}^{u}+1},  \notag
\\
&&  \label{sp.14}
\end{eqnarray}%
and the non-diagonal terms are given by%
\begin{eqnarray}
k_{12} &=&\beta _{12}\mathcal{G}^{(+)}(u),\quad k_{15}=\beta _{15}\mathcal{G}%
^{(+)}(u),\quad k_{16}=\beta _{16}\frac{\mathrm{e}^{u}+q^{3}}{1+q^{3}}%
\mathcal{G}^{(+)}(u),  \notag \\
k_{21} &=&\beta _{21}\mathcal{G}^{(+)}(u),\quad k_{25}=-\beta _{21}\frac{%
\beta _{16}}{\beta _{12}}\frac{\mathrm{e}^{u}+q^{3}}{1+q^{3}}\mathcal{G}%
^{(+)}(u),\quad k_{26}=\frac{\beta _{15}}{q^{3}}\mathrm{e}^{u}\mathcal{G}%
^{(+)}(u),  \notag \\
k_{51} &=&-\beta _{21}\frac{q^{4}\beta _{12}}{\beta _{15}}\mathcal{G}%
^{(+)}(u),\quad k_{52}=\beta _{21}\frac{q^{4}\beta _{12}\beta _{16}}{\beta
_{15}^{2}}\frac{\mathrm{e}^{u}+q^{3}}{1+q^{3}}\mathcal{G}^{(+)}(u),\quad
k_{56}=-q\beta _{12}\mathrm{e}^{u}\mathcal{G}^{(+)}(u),  \notag \\
k_{61} &=&-\beta _{21}^{2}\frac{q^{4}\beta _{16}}{\beta _{15}^{2}}\frac{%
\mathrm{e}^{u}+q^{3}}{1+q^{3}}\mathcal{G}^{(+)}(u),\quad k_{62}=-\beta _{21}%
\frac{q\beta _{12}}{\beta _{15}}\mathrm{e}^{u}\mathcal{G}^{(+)}(u),\quad
k_{65}=-q\beta _{21}\mathrm{e}^{u}\mathcal{G}^{(+)}(u),  \notag \\
&&  \label{sp.15}
\end{eqnarray}%
where%
\begin{equation}
\mathcal{G}^{(+)}(u)=\frac{1}{\beta _{16}}\left( \frac{1+q^{3}}{\mathrm{e}%
^{u}+q^{3}}\right) k_{16}\qquad \mathrm{and\qquad }\beta _{21}=-\frac{\beta
_{12}}{q^{3}}\frac{\beta _{15}^{2}}{\beta _{16}^{2}}+\frac{2}{(q+1)(q^{3}+1)}%
\frac{\beta _{15}}{\beta _{16}}.  \label{sp.16}
\end{equation}

For $n>3$ this type of solution follows the classification scheme presented
in section $4$.

\section{Conclusion}

In this work we have made a systematic study of the reflection equations for
the vertex models associated with C$_{n}^{(1)}$ ($sp(2n)$), D$_{n}^{(1)}$ ($%
o(2n)$) and A$_{2n-1}^{(2)}$ ($sl(n))$ affine Lie algebras. After looking
for $K$-matrix solutions with all entries different from zero we have found
two conjugated solutions for the $sl(2n)$ models and\ one solution for the $%
sp(2n)$ and also one solution for $o(2n)$ models. In both cases these
solutions have $n$ free parameters.

From the reduction procedure developed in section $4$ it has been possible
to find a second solution for the $sp(2n)$ and $o(2n)$ models. These new
solutions have the entries $k_{i,n}=k_{i,n+1}=k_{n,j}=k_{n,j+1}=0$ with $%
i,j\neq n,n+1$, $k_{n,n}=k_{n+1,n+1}\neq 0$ and $n-1$ free parameters. A
third solution with one free parameter has also been obtained from this
reduction procedure. Therefore, for $n>2$ we have found three types of
solutions for these models. For a given general solution there are $n-1$
possible reductions of this type but only the first and the last gives us
new solutions.

The special cases (with emphasis in the diagonal cases) were discussed
separately in the other sections of this paper.

\vspace{.5cm}{}

\textbf{Acknowledgment:} This work was supported in part by Funda\c{c}\~{a}o
de Amparo \`{a} Pesquisa do Estado de S\~{a}o Paulo-\textsl{FAPESP}-Brasil,
Conselho Nacional de Desenvolvimento-\textsl{CNPq}-Brasil and by Coordena%
\c{c}\~{a}o de Aperfei\c{c}oamento de Pessoal de N\'{\i}vel Superior--%
\textsl{CAPES}-Brasil.


\begin{thebibliography}{99}
\bibitem{Baxter} R.J. Baxter, \emph{Exactly Solved Models in Statistical
Mechanics} (Academic Press, 1982)

\bibitem{KIB} V.E. Korepin, A.G. Izergin and N.M. Bogoliubov, \emph{Quantum
Inverse Scattering Method and Correlation Functions}, Cambridge-1992.

\bibitem{ABR} E. Abdalla, M.C.B. Abdalla and K. Rothe, \emph{Nonperturbative
Methods in Two-Dimensional Quantum Field Theory }(Second Edition, World
Scientific, Singapore, 2001).

\bibitem{KR} P.P. Kulish and N.Yu. Reshetikhin, J. Sov. Math. 23 (1983) 2435

\bibitem{Jimbo} M. Jimbo, \emph{Commun. Math. Phys}. \textbf{102} (1986) 537

\bibitem{Cherednik} I. V. Cherednik, \emph{Theor. Math. Phys.} \textbf{61}
(1984) 977

\bibitem{Sklyanin} E. K. Sklyanin, \emph{J. Phys A: Math. Gen.} \textbf{21}
(1988) 2375

\bibitem{MN1} L. Mezincescu and R. I. Nepomechie, \emph{Int. J. Mod. Phys.} 
\textbf{A6} (1991) 5231 (addendum \textbf{A7} (1992) 5657)

\bibitem{Bazhanov} V.V. Bazhanov, \emph{Commun. Math. Phys.} \textbf{113}
(1987) 471

\bibitem{MN} L. Mezincescu and R. I. Nepomechie, \emph{Int. J. Mod. Phys.} 
\textbf{A13} (1998) 2747

\bibitem{ND} G.W. Delius and R.I. Nepomechie,\emph{\ J. Phys. A: Math. Gen.} 
\textbf{A35} (2002) L341

\bibitem{Nepo} R. I. Nepomechie, \emph{Lett. Math. Phys. }\textbf{62} (2002)
83

\bibitem{AACDFR} D. Arnaudon, J. Avan, N. Cramp\'{e}, A. Doikou, L. Frappat,
E. Ragoucy, \emph{Classification of reflection matrices related to (super)
Yangians}, math.QA/0304150

\bibitem{BCDR} P. Bowcock, E. Corrigan, P.E. Dorey and R.H. \ Rietdijk, 
\emph{Nucl. Phys.} \textbf{B445} (1995) 469

\bibitem{Batch1} M. T. Batchelor and C. M. Yung, \emph{Phys. Rev. Lett.} 
\textbf{74} (1995) 2026

\bibitem{Batch2} M. T. Batchelor, V. Fridkin, A. Kuniba and Y. K. Zhou, 
\emph{Phys. Lett.} \textbf{B376} (1996) 266

\bibitem{RE1} H. J. de Vega and A. Gonz\'{a}lez-Ruiz, \emph{J. Phys. A:
Math. Gen.} \textbf{27} (1994) 6129

\bibitem{RE2} C. X. Liu, G. X. Ju, S.K. Wang, K. Wu, \emph{J. Phys. A: Math.
Gen.} \textbf{32} (1999) 3505

\bibitem{RE3} L. Mezincescu and R. I. Nepomechie, \emph{Int. J. Mod. Phys.} 
\textbf{A6} (1991) 5231

\bibitem{RE4} T. Inami, S. Odake, Y. Z. Zhang, \emph{Nucl. Phys}. \textbf{%
B470} (1996) 419

\bibitem{RE5} J. D. Kim, \emph{Boundary K-matrix for the quantum
Mikhailov-Shabat model}, hep-th/9412192

\bibitem{RE6} A. Lima-Santos, \emph{Nucl. Phys.} \textbf{B558} [PM] (1999)
637

\bibitem{RE7} H. J. de Vega and A. Gonz\'{a}lez-Ruiz, \emph{J. Phys. A:
Math. Gen.} \textbf{26} (1993) L519

\bibitem{RE8} J. Abad and M. Rios, Phys. Lett. B352 (1995) 92

\bibitem{RE9} A. Lima-Santos, \emph{Nucl. Phys.} \textbf{B644} [FS] (2002)
568

\bibitem{RE10} M. J. Martins, X. W. Guan, \emph{Nucl. Phys.} \textbf{B583}
(2000) 721

\bibitem{RE11} A. Lima-Santos, \emph{Nucl. Phys.} \textbf{B612} [FS] (2001)
446

\bibitem{RE12} A. Lima-Santos, \emph{Nucl. Phys.} \textbf{B654} [FS] (2003)
466

\bibitem{Pearce} R. \ E. Behrend and P. A. Pearce, \emph{Int. J. Mod. Phys.} 
\textbf{B11} (1997) 2833

\bibitem{Ahn} C. Ahn and C. K. You, \emph{J. Phys. A: Math. Gen.} \textbf{31}
(1998) 2109

\bibitem{MA} M. J. Martins, \emph{Quantum group Bethe Ansatz solution of the
A}$_{2n-1}^{(2)}$\emph{\ open spin chain}., to be published.
\end{thebibliography}
\end{document}